\documentclass[9pt,twocolumn,twoside]{pnas-new}

\templatetype{pnasresearcharticle} 

\usepackage{graphicx}
\usepackage{subcaption}
\usepackage{xcolor}
\usepackage{longtable} 
\usepackage{booktabs}  
\usepackage{siunitx}
\usepackage{caption}   
\usepackage{hyperref} 
\usepackage{booktabs}
\usepackage{array}
\usepackage{tabularx}

\begin{document}

\title{The Tilted Playing Field for Women in Science}

\author[a,b,1,2]{Casandra Rusti}
\author[c,b]{Hussain Hussain}
\author[a,b]{Kian Ahrabian}
\author[a,b]{Jay Pujara}
\author[d]{Allon G.\ Percus}
\author[e]{Buddhika Nettasinghe}
\author[f,a,b,1]{Kristina Lerman}

\affil[a]{University of Southern California}
\affil[b]{Information Sciences Institute (ISI)}
\affil[c]{Technische Universität Graz}
\affil[d]{Claremont Graduate University}
\affil[e]{The University of Iowa}
\affil[f]{Indiana University Bloomington}


\leadauthor{Rusti}


\significancestatement{Success in science is strongly associated with institutional prestige, but whether prestige benefits researchers equally remains poorly understood. Using large-scale data on papers and co-authorship, we show that affiliation with higher-ranked institutions is associated with higher levels of productivity and collaboration – but not equally for all researchers. Gender differences emerge and widen at higher levels of achievement and coincide with differences in collaboration network structure and the distribution of co-author ties across institutional hierarchies. These findings suggest that inequality in science reflects not only individual performance, but also how opportunities are structured through prestige and networks, revealing a tilted playing field that shapes who succeeds and who does not.}

\authorcontributions{K.L., B.N. conceptualized the study;
C.R., B.N., A.P., J.P., and K.L. developed analysis methodology;
C.R., H.H., and K.A. analyzed the data;
C.R., K.L. wrote the initial draft;
all authors reviewed and edited the manuscript.}
\authordeclaration{The authors declare no competing interest.}
\equalauthors{\textsuperscript{1}C.R. and K.L. contributed equally to this work.}
\correspondingauthor{\textsuperscript{2}To whom correspondence should be addressed. E-mail: rusti@usc.edu}


\begin{abstract}
Institutional prestige shapes access to resources, visibility, and collaboration opportunities in science. Yet whether prestige benefits researchers equally, and how it relates to differences in scientific productivity and collaboration, remains unclear. Here, we quantify prestige advantage as the relative likelihood that researchers at higher-ranked institutions have more collaborators and produce more high-impact papers compared to their lower-ranked peers. Analyzing nearly 5 million papers by 6.5 million authors across more than 65,000 institutions, we present a distributional, tail-sensitive framework to compare prestige advantage across groups. We find that the association between prestige and scientific achievement differs systematically by gender. While both men and women benefit from prestige, the returns are not gender-neutral: women experience comparable advantages only at the most elite institutions, whereas men retain persistent advantages across the broader hierarchy, with disparities widening at higher levels of achievement. Prestige advantage also grows nonlinearly, disproportionately benefiting authors at the most elite institutions. These differences align with collaboration patterns: women’s networks are more locally clustered and focused on their own institution, while men collaborate more broadly across institutional strata. Together, these findings reveal a tilted playing field in science: one where prestige amplifies success unevenly and network structure shapes who can access its benefits.
\end{abstract}

\dates{Preprint. Under review.}
\doi{\url{www.pnas.org/cgi/doi/10.1073/pnas.XXXXXXXXXX}}

\maketitle
\thispagestyle{firststyle}
\ifthenelse{\boolean{shortarticle}}{\ifthenelse{\boolean{singlecolumn}}{\abscontentformatted}{\abscontent}}{}

\firstpage[9]{4}

\dropcap{S}cientific careers unfold within a stratified system of institutions in which prestige shapes access to resources, collaborations, and visibility. Researchers at elite institutions consistently outperform counterparts with less prestigious affiliations on key metrics of academic success, including publications,  citations, research funding and awards~\cite{clauset2015systematic,way2019productivity,bol2018matthew}. 
This pattern is often interpreted as the outcome of a meritocratic process in which talent rises, elite institutions attract the best minds and provide the conditions necessary for further success~\cite{strevens2006role,xie2014undemocracy}. Prestige then functions as a self-reinforcing cycle:  achievement places talented researchers at elite institutions, which in turn confer labor, visibility, and resource advantages that further amplify their success~\cite{burris2004academic,way2016gender,way2019productivity,allison1990departmental,heckman2020publishing,bol2018matthew,zhang2022labor,burghardt2021emergence}. 

However, a growing body of evidence suggests that science is not purely meritocratic. Women scholars receive less recognition than equally productive male peers~\cite{dion2018gendered,lerman2022gendered} and are evaluated less favorably in funding decisions~\cite{witteman2019gender}. Gender disparities also appear in institutional dynamics: while elite universities hire more women than expected~\cite{way2016gender}, women are more likely to receive tenure in lower-ranked departments~\cite{weisshaar2017publish} and exhibit higher attrition rates than men~\cite{huang2020historical,spoon2023gender}.  
These findings raise a question: does institutional prestige amplify success equally across genders, or does it differentially shape long-term career trajectories? The answer has important implications for both the fairness of the scientific enterprise and its ability to  identify talent.


Here, we examine how institutional prestige and gender jointly shape scientific success. We analyze authorship across five million scientific papers published between 1980 and 2024 in 100 high-impact venues spanning science, engineering, and medicine, providing a consistent baseline of research quality. From this dataset, we identify the gender and institutional affiliations of approximately 4.3 million authors, enabling a large-scale analysis of how prestige interacts with gender. 


We find that institutional prestige is strongly associated with higher productivity and broader collaboration for all researchers. However, the benefits of prestige are not distributed equally. At the most elite institutions, women experience advantages comparable to their male counterparts. But beyond the top 100, this advantage erodes sharply: women at lower-ranked institutions receive little measurable benefit from prestige, while men continue to benefit across all institutional tiers. Because 88\% of researchers are located outside the top 100 institutions, this asymmetry affects the vast majority of scientists. The playing field, it turns out, is tilted. 

We identify a potential explanation for this asymmetry. Men form diverse collaborations that cross prestige boundaries, reaching upward from lower-ranked institutions to create higher-prestige connections. Women, by contrast, collaborate more within their own or similarly ranked institutions. This difference in collaboration architecture---rather than differences in research quality---drives the observed disparity, limiting women's prospects for career mobility. 


\section{Results}

We analyze nearly 5 million papers published between 1980 and 2024 in the 100 highest-impact academic venues, as ranked by Google Scholar~\cite{google_scholar_top_publications} (see Methods). These include interdisciplinary journals such as Nature, Science, and JAMA, as well as leading disciplinary venues such as NeurIPS, Cell, Chemical Reviews, and Energy \& Environmental Science. Because these venues apply highly selective editorial and peer-review standards, they offer a consistent baseline of scientific merit, allowing us to disentangle the effects of prestige from underlying differences in research quality. 
The resulting dataset comprises over 6.5 million authors affiliated with more than 65,000 institutions worldwide. To capture the hierarchical structure of science,  we map author affiliations to the top 2000 universities in the 2025 Times Higher Education World University Rankings (THE)~\cite{the_rankings_2025}. 

We infer author gender based on names (see Methods), identifying gender for approximately 4.3 million authors. While women are a minority in our dataset, their representation increases over time, rising from fewer than 20 women per 100 men in 1980 to approximately 55 women per 100 men by 2022, corresponding to roughly one-third of authors in recent years (SI Fig.~\ref{fig:women_by_year}).

\begin{figure*}[t]
\centering
\begin{tabular}{cc}
    \includegraphics[width=0.4\linewidth]{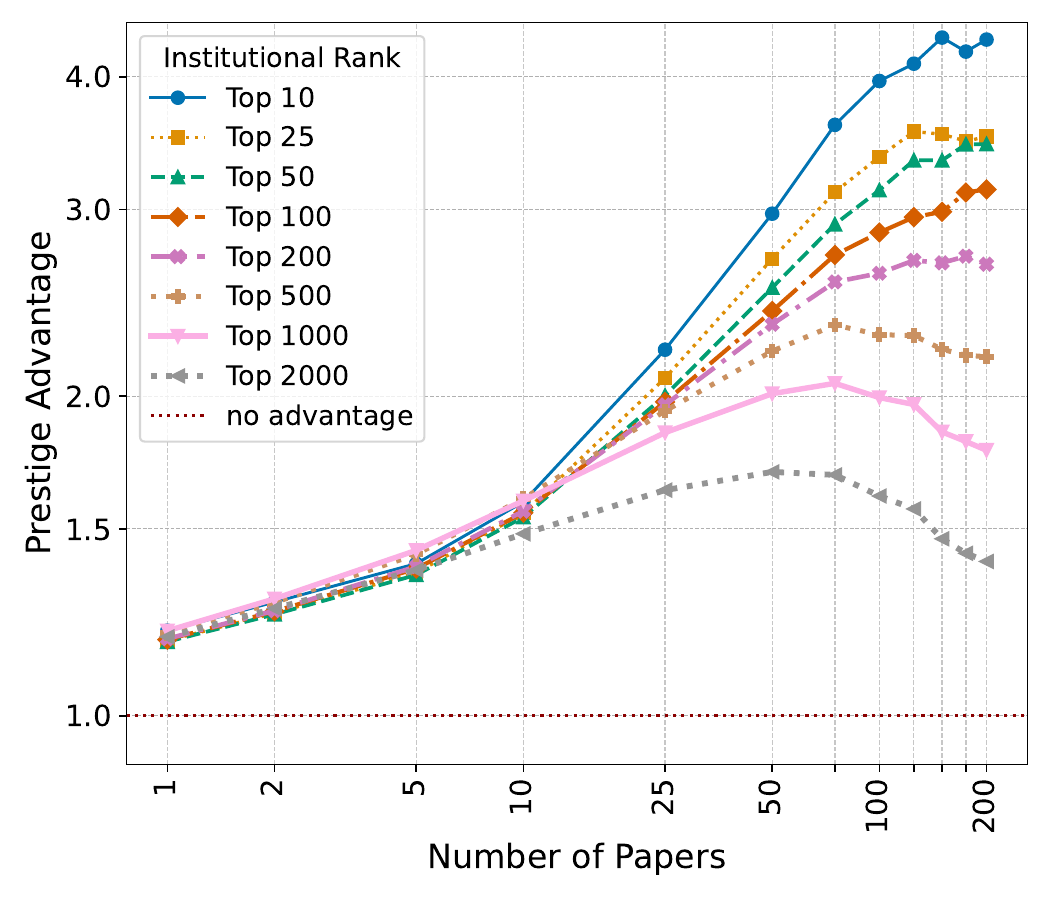}
&    \includegraphics[width=0.4\linewidth]{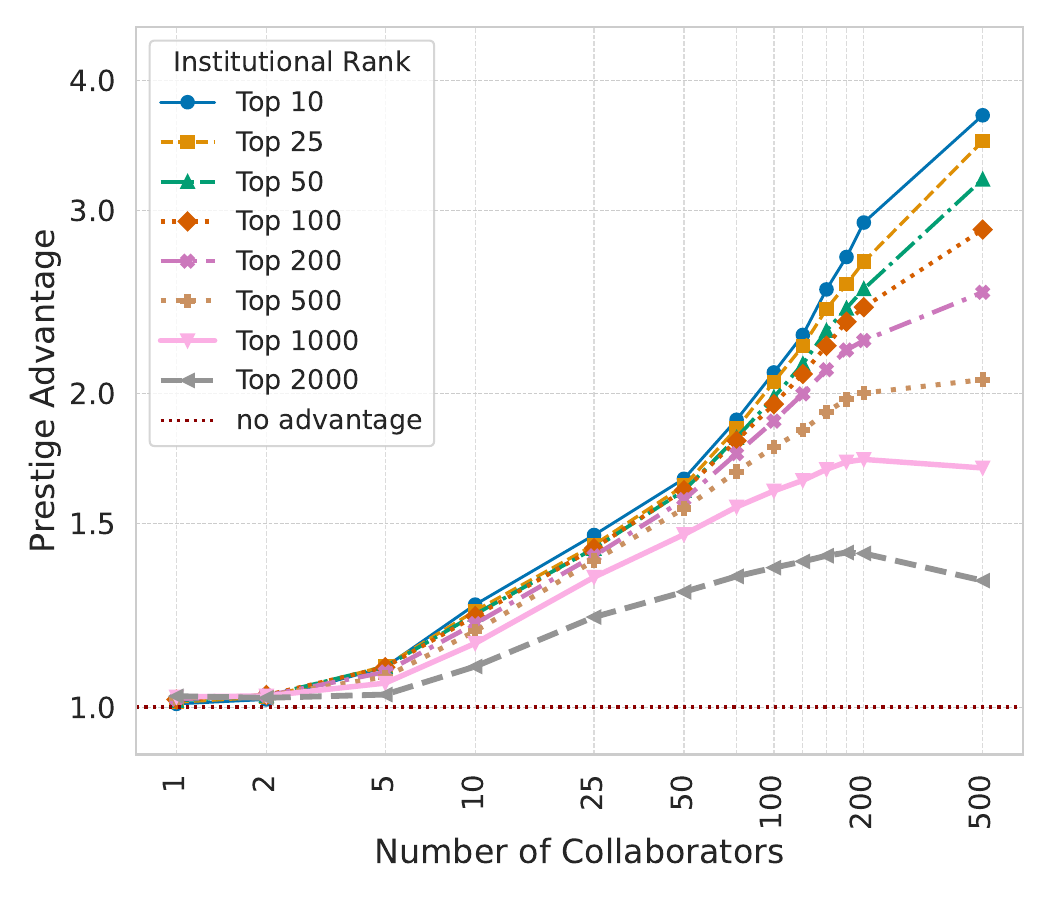}
\\
(a) & (b)
\end{tabular}
    \caption{\textbf{Prestige advantage.}
    The figure shows the stratification of prestige advantage across institutional ranks. Prestige advantage measures how much more likely it is to find a highly productive or well-connected author at a top-$k$ institution relative to the rest of the academic system. Specifically, it plots the ratio of probabilities that an author affiliated with a top-$k$ institution in the \textit{Times Higher Education} (THE) 2025 World University Rankings exceeds a given threshold $x$ of (a) papers or (b) collaborators, relative to the corresponding probability for authors not affiliated with a top-$k$ institution. Values greater than one indicate that prolific or highly collaborative authors are overrepresented at higher-ranked institutions. Each line corresponds to a different institutional prestige level $k$ (see legend), revealing a clear stratification.}
    \label{fig:prestige-THE-ranking}
\end{figure*}

\subsection{Prestige Advantage}

To quantify how institutional prestige is associated with differences in scientific productivity and collaboration,  we examine the distribution of papers (or collaborators) per author across prestige tiers, using the complementary cumulative distribution function (CCDF). The CCDF gives the probability $P(X>x)$ that a randomly selected author has more than $x$ papers (or collaborators). The CCDF is particularly well suited here because academic productivity is highly skewed: most authors have few papers in high-impact venues, while a small fraction are highly prolific (SI Fig.~\ref{fig:prestige-THE-CCDF}). The CCDF allows us to capture this tail behavior, where prestige differences are most consequential,  without imposing assumptions on the shape of the distribution.

The CCDFs of papers and collaborators per author are not only highly skewed but also stratified by prestige. To compare productivity across institutional tiers, we define \textit{prestige advantage} as the ratio of CCDFs between authors inside and outside a given prestige class (Eq.~\ref{eq:prestige-adv}). Values greater than one indicate that affiliation with higher-ranked institutions is associated with a greater probability of having more than $x$ papers (or collaborators). This formulation has a natural interpretation: if $A(10, 100)=3$, authors at top-10 institutions are three times more likely to have published over 100 papers in high-impact venues than authors outside that group.

The prestige advantage (Fig.~\ref{fig:prestige-THE-ranking}) shows three clear trends.  First, prestige induces a clear hierarchy of advantage: authors at more highly ranked institutions are consistently more likely to exceed any given threshold. 
Second, prestige advantage grows with the threshold: authors at top-10 institutions are modestly more likely to exceed lower thresholds (e.g., 5 papers) than their peers outside the top-10 institutions, but four times more likely to have produced more than 100 papers, indicating amplification in the upper tail of the distribution.
Third, this amplification is strongest among the most elite institutions, concentrating highly prolific authors within a small set of top-ranked affiliations, while prestige advantage plateaus or even declines among lower-ranked groups. 
Both papers and collaborators exhibit these patterns (Fig.~\ref{fig:prestige-THE-ranking}a,b), though prestige advantage is slightly smaller and less sharply stratified for collaborations than for publications. Together, these show that institutional prestige systematically shapes the probability of reaching high levels of scientific productivity and collaboration in top venues.

\begin{figure*}[t]
\centering
\begin{tabular}{>{\centering\arraybackslash}p{0.68\linewidth}
                >{\centering\arraybackslash}p{0.32\linewidth}}
    \includegraphics[width=\linewidth]{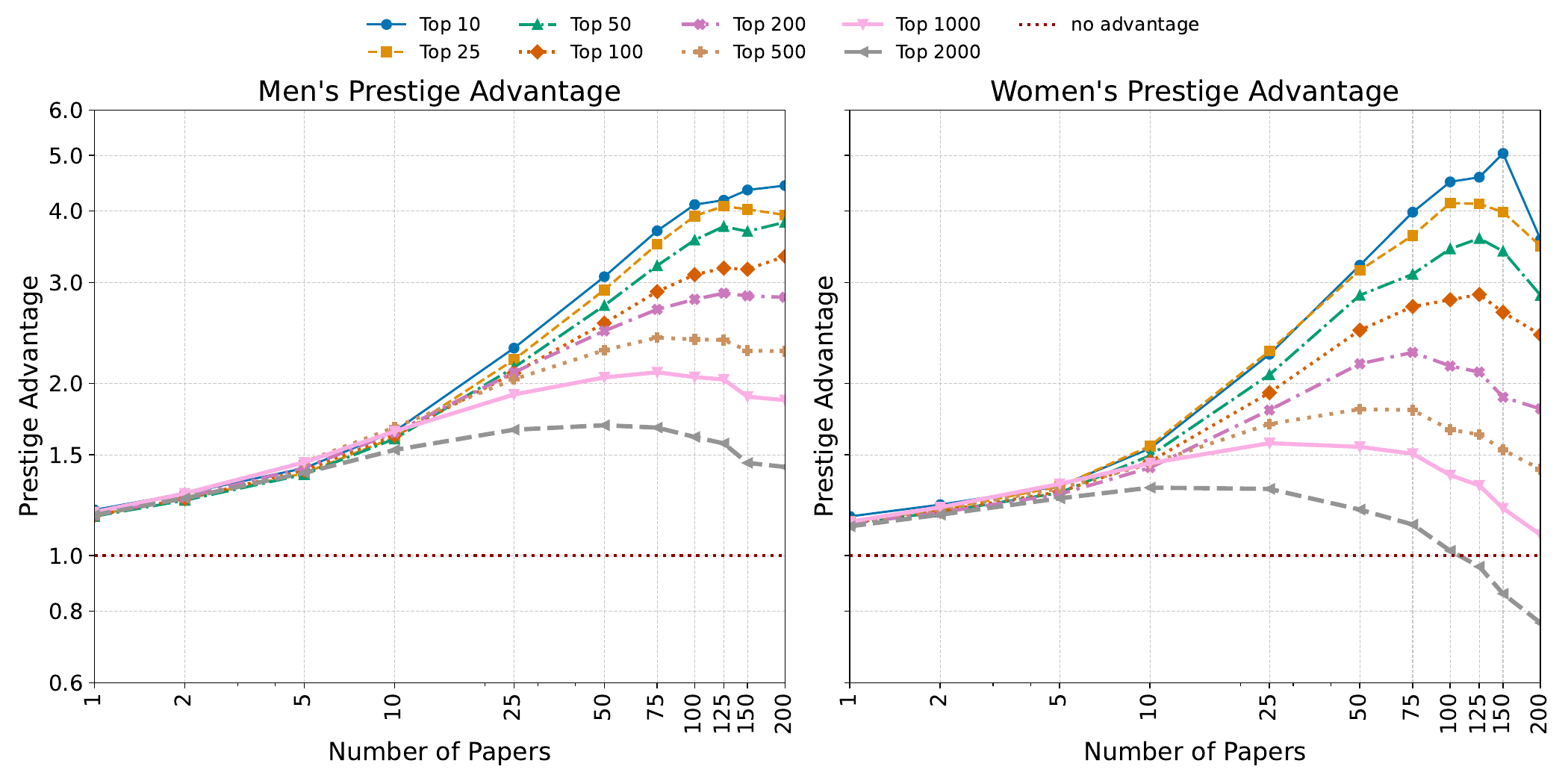} &
    \includegraphics[width=\linewidth]{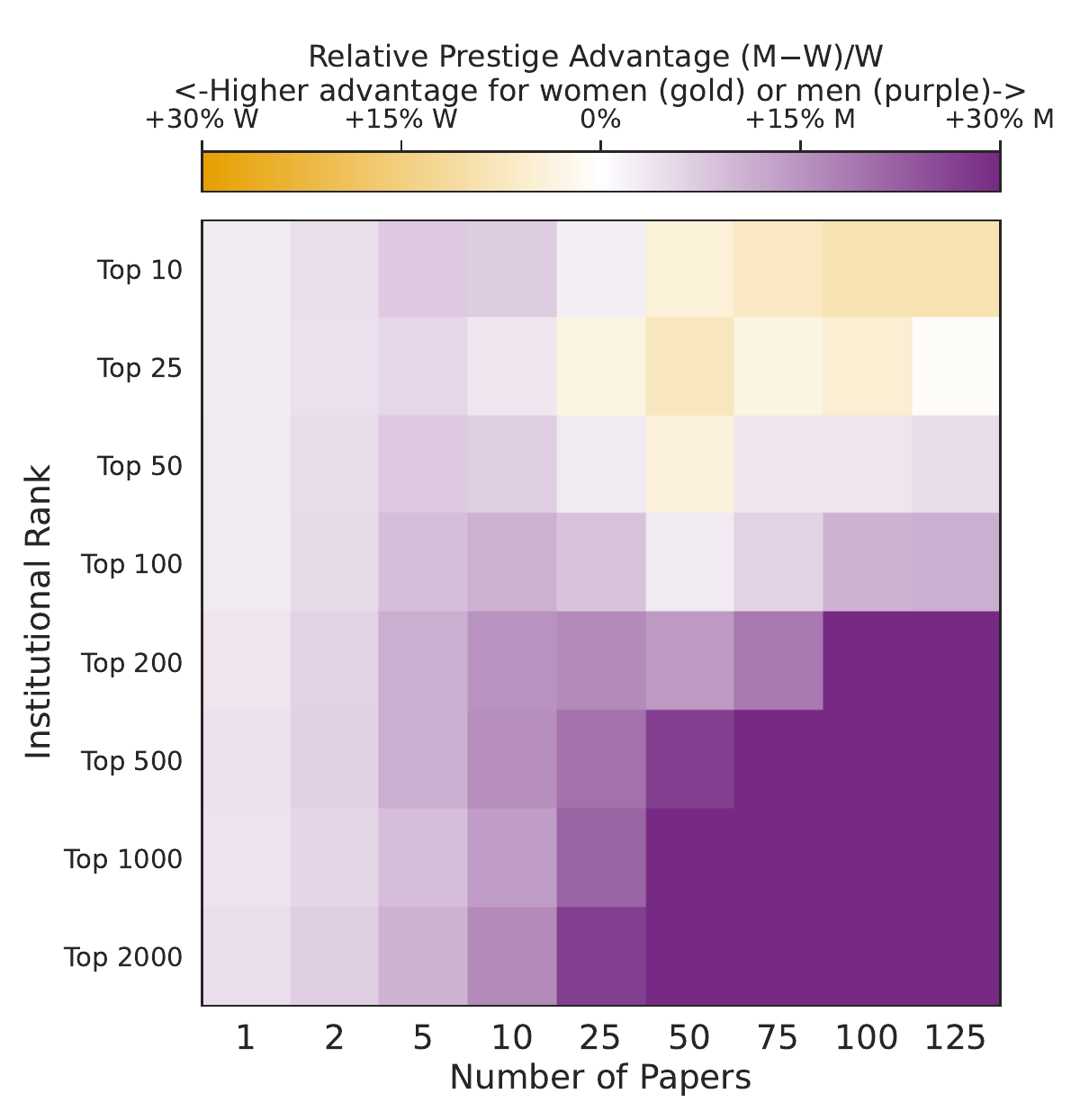} \\
    (a) & (b) \\
    \includegraphics[width=\linewidth]{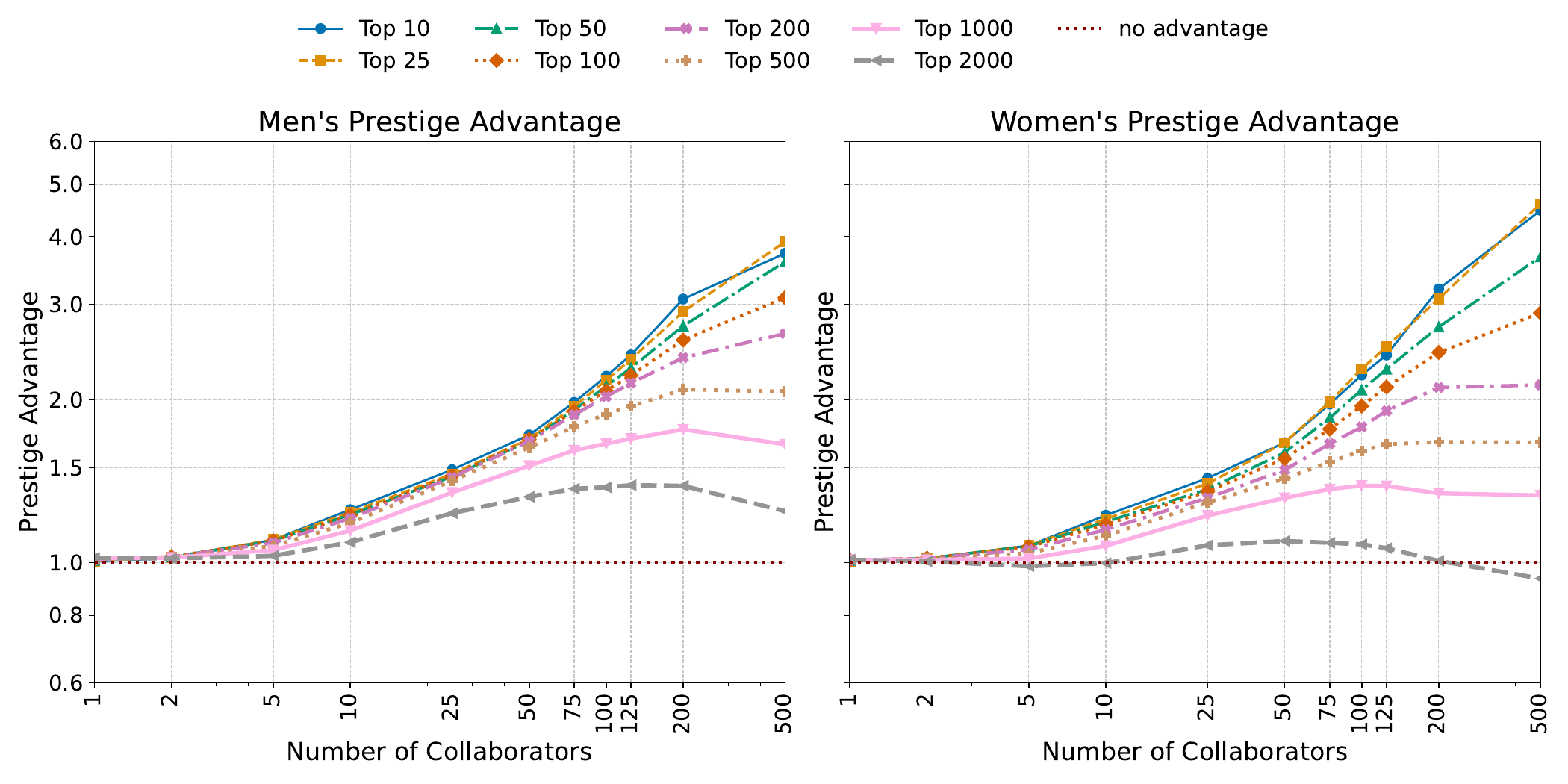} &
    \includegraphics[width=\linewidth]{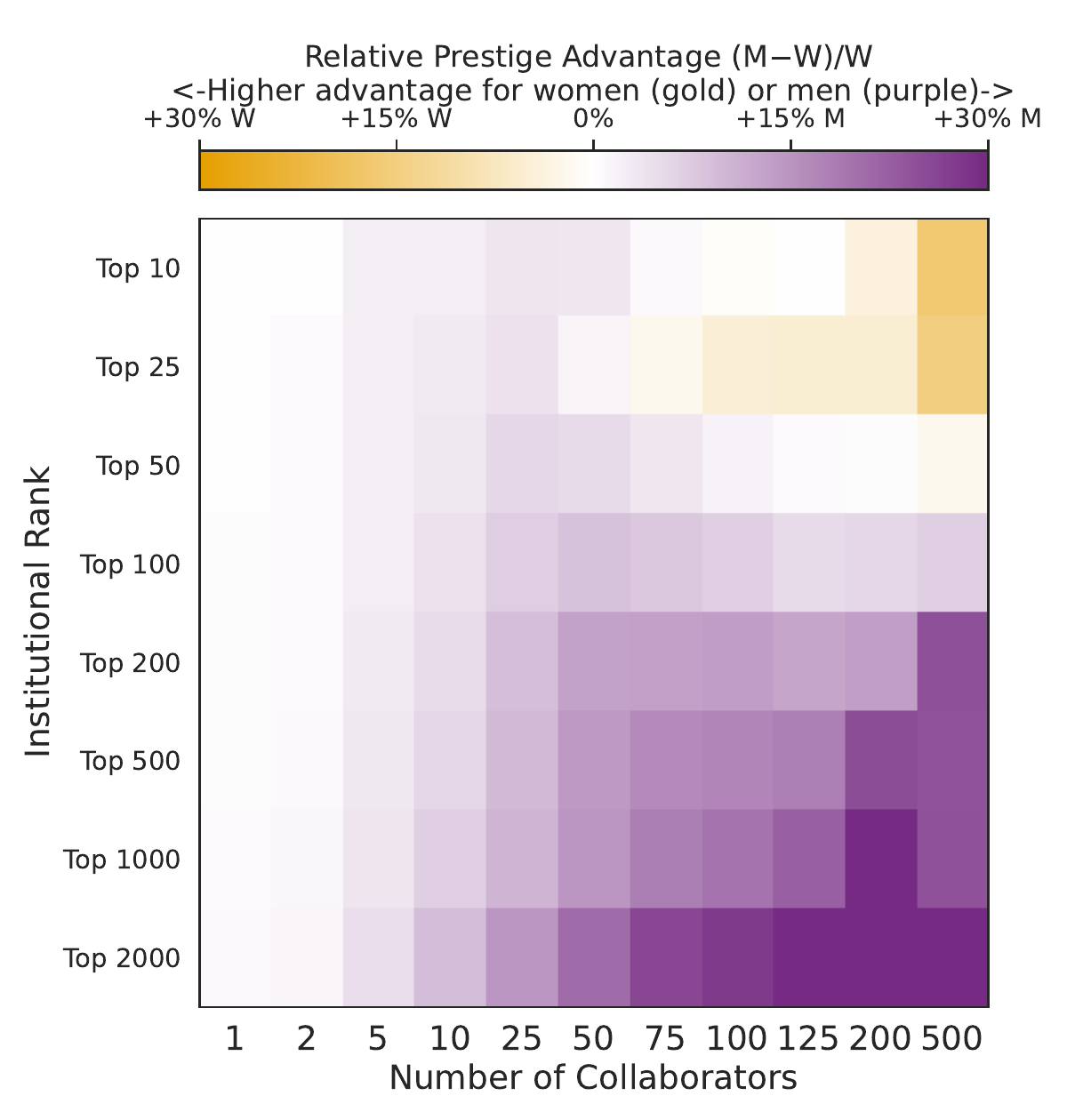} \\
    (c) & (d)
\end{tabular}
    \caption{\textbf{Gender differences in prestige advantage.}
    Prestige advantage is shown separately for men and women as a function of institutional rank. Panels (a) and (c) report prestige advantage for papers and collaborators, respectively. Panels (b) and (d) visualize gender gap in prestige advantage using the relative difference $(M - W)/W$, where $M$ and $W$ denote men’s and women’s prestige advantage evaluated at the same threshold. Values greater than zero indicate that institutional prestige is more strongly associated with greater advantage for men than for women, while negative values indicate the opposite. Institutional ranking is based on the \textit{Times Higher Education} (THE) 2025 World University Rankings. To maintain robust sample sizes in the upper tail of the productivity distribution, the visualization is restricted to thresholds up to $x = 125$ papers and $x = 500$ collaborators. At these respective thresholds, the smallest group corresponds to authors affiliated with top-10 institutions and exceeding 125 papers (293 men and 51 women) or exceeding 500 collaborators papers (202 men and 39 women).}
\label{fig:prestige-gender-THE-ranking}
\end{figure*}

\subsection{Gender Differences in Prestige Advantage}

Next we ask whether this amplification operates equally for men and women by computing prestige advantage separately for each gender. 
By comparing women only to other women, we isolate the effects of institutional affiliation from broader structural differences, such as women having shorter careers~\cite{huang2020historical} and smaller collaboration networks~\cite{jadidi2017gender} than men. 

Prestige advantage is present for both men and women: affiliation with higher-ranked institutions increases the likelihood of exceeding any given threshold in both papers and collaborations (Fig.~\ref{fig:prestige-gender-THE-ranking}a,c). Yet the magnitude and structure of this advantage differ systematically by gender.

For papers, prestige advantage grows with productivity for both groups, but the two diverge as the threshold rises (Fig.~\ref{fig:prestige-gender-THE-ranking}a). Among the most elite institutions, women's prestige advantage is comparable to that of men at moderate thresholds and exceeds it at higher productivity levels. Below the top-100, however, the picture reverses: women's prestige advantage does not grow substantially or saturates with institutional rank. The result is a wider dispersion of curves for women than for men: the gap between elite and non-elite institutions is larger for women, meaning that institutional rank matters more -- not less -- for women's careers.
These differences cannot be explained by career length, as there is no systematic variation in career length with institutional prestige for either gender (SI Fig.~\ref{fig:career_length_and_table}).

To quantify this asymmetry, we compute the relative gender gap in prestige advantage across thresholds and institutional ranks (Eq.~\ref{eq:gender-delta}). Figure~\ref{fig:prestige-gender-THE-ranking}b shows the heat map of this quantity. The pattern is striking: at the highest ranked institutions and highest productivity thresholds, women exhibit comparable or slightly greater prestige advantage than men. Across the broader institutional landscape, however, men consistently enjoy larger prestige advantages, with disparities widening at higher productivity thresholds and lower institutional ranks.

Collaboration patterns tell a consistent story (Fig.~\ref{fig:prestige-gender-THE-ranking}d). Women benefit from institutional prestige in building collaborations, but their relative advantage remains consistently below that of men across most of the distribution.

Together, these results reveal that prestige advantage is not gender-neutral: both men and women benefit from institutional prestige, but women match or exceed men’s advantage only at the very top, while men retain a relative advantage across the rest of the institutional hierarchy. This is the tilted playing field.

\subsection{Collaboration Network Structure Differs by Gender}

Finally, we ask whether gender differences in prestige advantage are reflected in collaboration structure. We examine two complementary aspects of co-authorship network topology: local clustering in the full network (Approach A) and clustering within prestige-restricted subgraphs (Approach B; see Methods).

\begin{figure}[htbp]
    \centering
    \includegraphics[width=\linewidth]{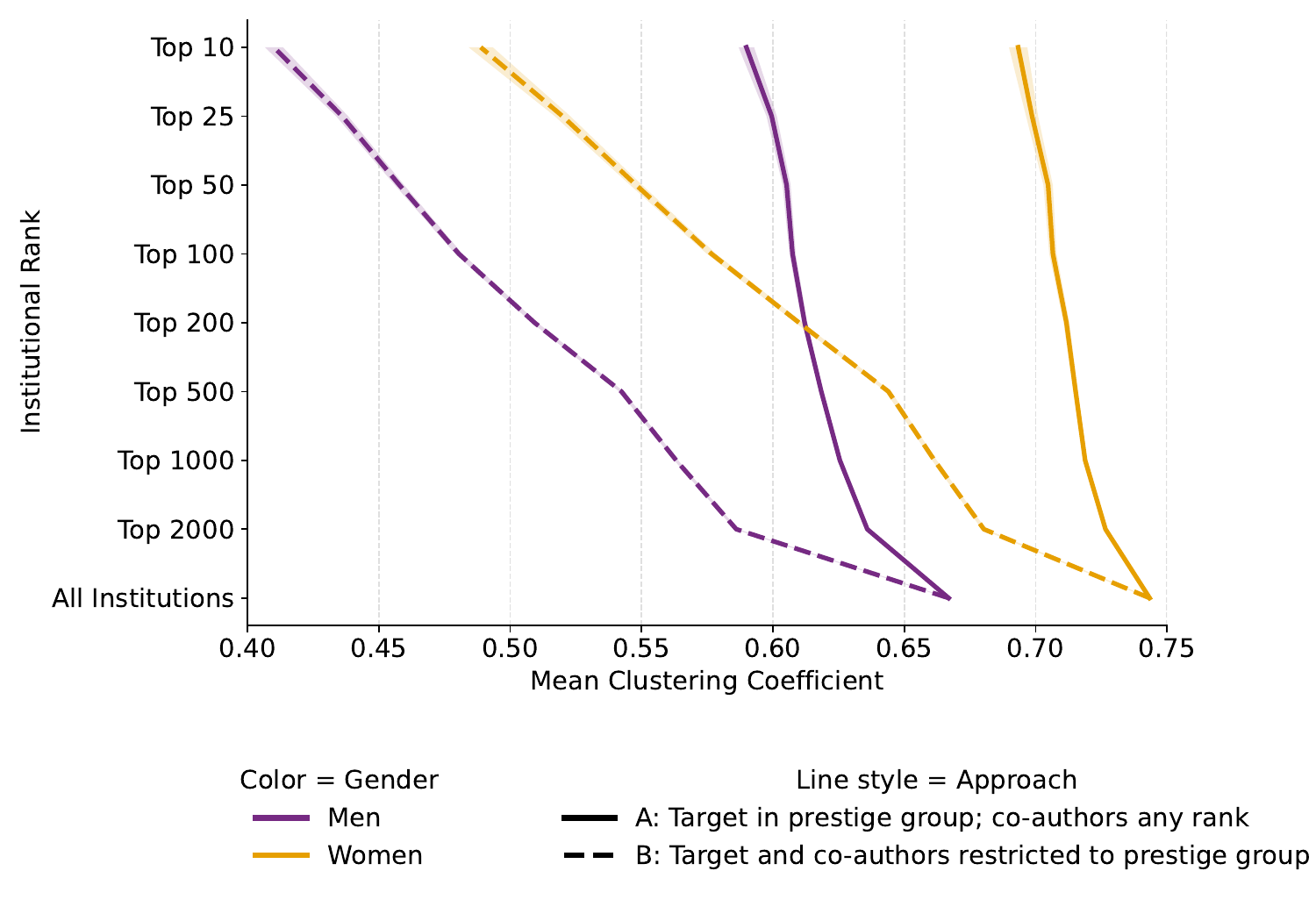}
    \caption{\textbf{Gender differences in collaboration network clustering across institutional rank.}
    Mean clustering coefficients of authors’ collaboration networks are shown as a function of the institutional rank used to define the prestige group. Solid lines (Approach A) correspond to clustering computed on the full collaboration network, where authors are grouped by institutional rank, while dashed lines (Approach B) restrict clustering to subgraphs in which both the focal author and collaborators belong to the same institutional rank group. Colors indicate gender. In all cases, women are in more highly clustered networks than men. Clustering increases as the institutional rank threshold is relaxed for all groups. Shaded bands represent 95\% confidence intervals around the mean clustering coefficients, which are narrow and largely overlap with the plotted lines. Despite differences in mean values, clustering coefficients are highly dispersed: the 75th percentile is consistently equal to 1 across genders, institutional rank thresholds, and both approaches, while the 25th percentile is near zero for Approach B and approximately 0.2 below the mean for both men and women for Approach A. Institutional ranking is based on the \textit{Times Higher Education} 2025 World University Rankings.}
    \label{fig:cluesting_gender_prestige}
\end{figure}

Across all institutional rank thresholds, women are embedded within more locally dense co-authorship networks, as indicated by higher clustering coefficients, than men (Fig.~\ref{fig:cluesting_gender_prestige}). This holds whether clustering is computed on the full co-authorship network or restricted to collaborators within the same institutional rank group. Restricting to within-rank collaborations reveals an important structural distinction: clustering coefficients decline substantially for all groups, particularly among higher-ranked institutions, indicating that the collaborators of elite researchers frequently connect across institutional boundaries rather than forming closed loops within the same prestige tier. Despite this overall decline, women maintain higher clustering across all rank thresholds, suggesting that gendered differences in collaboration topology are not simply a byproduct of institutional affiliation, but reflect broader patterns in co-authorship formation.


\begin{figure}[htbp]
    \centering
    \includegraphics[width=\linewidth]{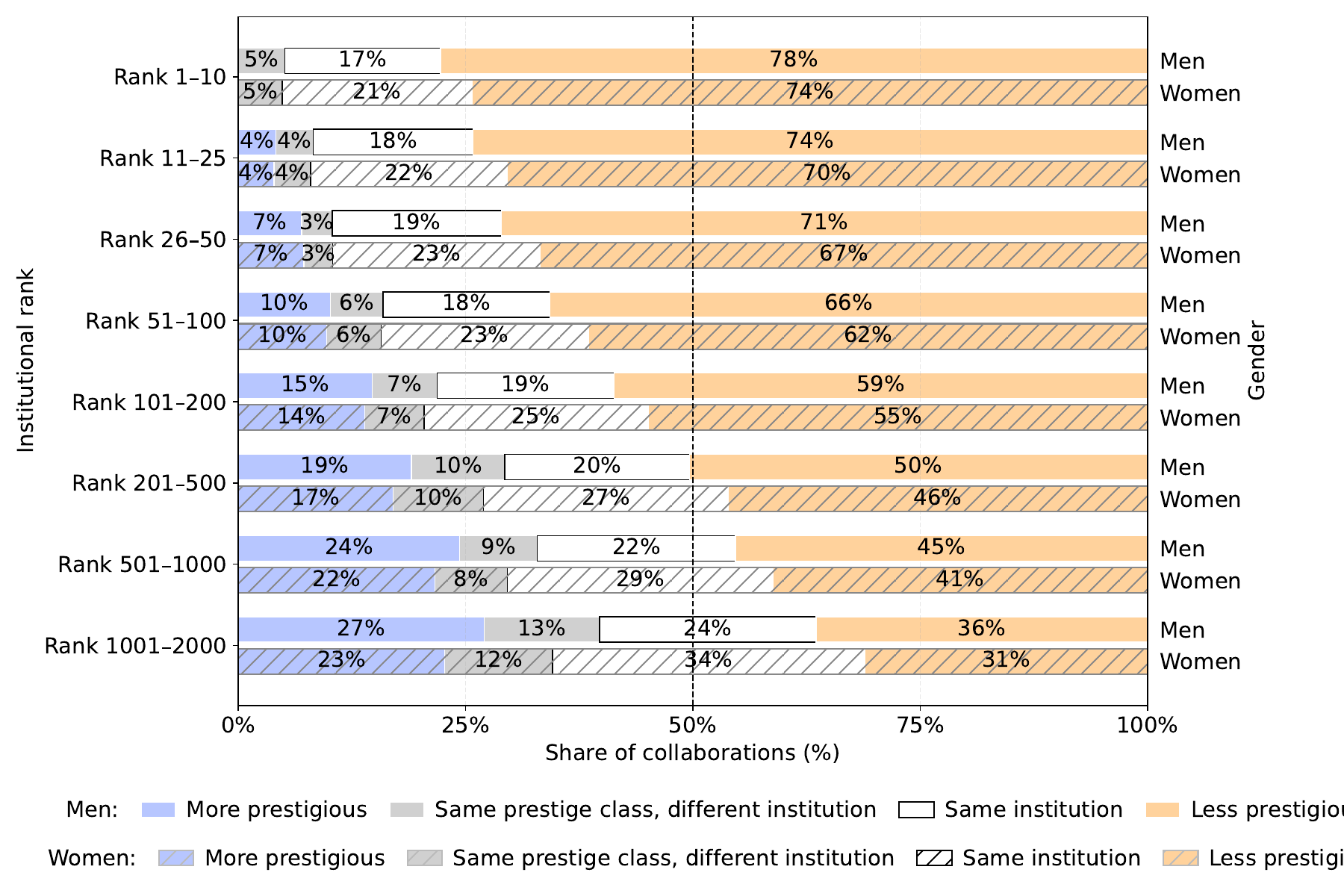}
    \caption{\textbf{Collaborator prestige by institutional rank and gender.}
    The distribution of collaborator institutional prestige for focal authors split by gender and institutional rank. For each group, collaborators are partitioned into four mutually exclusive categories: collaborators at more prestigious institutions, collaborators in the same institutional rank class, collaborators within the same institution (internal), and collaborators at less prestigious institutions. Solid bars represent men; hatched bars represent women. Across prestige classes, women have a larger share of internal collaborations, whereas men allocate a larger fraction of ties to collaborators outside their own institutional rank class. Moving from the most highly ranked institutions toward lower-ranked classes, collaborations become increasingly concentrated within the same institution or rank class for both genders (see Methods).}
    \label{fig:Collaboration_prestige_mix}
\end{figure}


We find systematic gender differences in the composition of collaboration ties (Fig.~\ref{fig:Collaboration_prestige_mix}). Women's collaborations are more concentrated within their own institution, whereas men allocate a larger share of ties to collaborators outside their rank. This pattern persists across the institutional hierarchy: as prestige decreases, collaborations become increasingly localized for both genders, but women’s networks remain consistently more rank-constrained, while men maintain broader cross-rank connections at every level.

\section{Discussion}

Science is not a level playing field. Resources, visibility, and opportunity are unevenly distributed across the institutional landscape, and our results show that these inequalities systematically benefit some groups but not others. Institutional prestige acts as a productivity amplifier, one whose effects are strongest in the upper echelons of institutional prestige, where the most prolific and well-connected researchers reside.
Yet this amplification is not gender-neutral. Women benefit from institutional prestige primarily at the very top of the institutional hierarchy. Below the top-100 institutions, this advantage weakens sharply, approaching parity with non-elite and unranked institutions. Men, by contrast, retain a meaningful  advantage across the full spectrum of prestige. This results in a steeper prestige gradient for women, reflecting a more stratified system in which the gap between elite and non-elite institutions is larger and the pathways for upward mobility narrower~\cite{jadidi2017gender}.

What drives this asymmetry? Our results point to the structure of collaboration networks as a potential mechanism. Women’s collaboration networks are more locally clustered---a pattern consistent with prior work on gender differences in brokerage and bridging ties~\cite{jadidi2017gender,burt2005brokerage}--but also more confined within institutional boundaries. Men, by contrast, maintain broader, cross-rank connections, enabling researchers at lower-prestige institutions to access the resources and visibility of higher-prestige networks. This facilitates the accumulation of social capital and, in turn, supports upward mobility. Women’s more rank-constrained collaboration patterns limit access to such capital, restricting the pathways through which institutional mobility becomes possible.

Paradoxically, this same tendency toward within-institution collaboration may explain women's prestige advantage at the most elite institutions. By concentrating collaborations locally, women at top-ranked universities can fully leverage the labor and resource advantages of these environments in the form of larger research groups, more graduate students, and greater collaborative infrastructure~\cite{zhang2022labor}. At elite institutions, internal collaborations are productive. At lower-ranked institutions, however, the same strategy yields diminishing returns, as the local environment offers fewer resources to sustain high productivity. The result is a self-reinforcing loop: collaboration patterns that are advantageous at the top become limiting further down the hierarchy, amplifying stratification and constraining mobility.

These patterns are unlikely to arise solely from individual preferences. Collaboration networks are shaped not only by individual choices but also by the opportunities that structure professional interactions. Early access to mentorship, visibility, and brokerage positions connecting otherwise distant parts of the network plays a critical role in long-term success~\cite{bachmann2024brokerage}. If women are systematically underrepresented in such positions, the resulting disadvantages compound over time~\cite{rossiter1993matthew}.

We also note that differences in career length and attrition contribute to the observed distributions. Women leave academia at higher rates and earlier career stages~\cite{huang2020historical}, affecting the composition of senior researchers. We also observe this in our data (Supplementary Fig.~\ref{fig:career_length_and_table}). However, the gender asymmetry in prestige advantage persists even when comparing authors at similar productivity levels, indicating that attrition alone cannot explain the tilted playing field.

Taken together, our findings provide a unified account of how inequality persists in science. The prestige system is more stratified for women than for men, and this stratification both reflects and reinforces limited mobility. Women’s rank-constrained collaboration can restrict access to the resources and social capital that flow through higher-prestige institutions, limiting the benefits of collaboration and narrowing pathways for advancement. Men, in contrast, can leverage prestige across institutional tiers through broader, cross-rank collaboration networks. Addressing this asymmetry requires more than increasing representation at elite institutions. It requires expanding the pathways through which collaboration, resources, and opportunity circulate across the full institutional hierarchy.

Several limitations should be considered. First, our analysis is observational and does not establish causal relationships between institutional prestige, collaboration patterns, and the accumulation of papers and collaborators. Second, gender is inferred from names and may be subject to classification error. Third, by focusing on high-impact venues, we capture a specific segment of scientific activity, and our findings may not generalize to all disciplines or publication contexts. Finally, institutional rankings provide only an approximate measure of prestige and may not fully capture all dimensions of academic status.

In summary, our results show that institutional prestige plays a central role in shaping scientific careers, but that its benefits are unevenly distributed. By linking prestige advantage to the structure of collaboration networks, we reveal how inequality is both embedded and sustained within the scientific enterprise.

\matmethods{
\paragraph{Bibliographic Corpus.}
We use OpenAlex bulk snapshot of 21 July 2025~\citep{priem2022openalex}, restricting to papers published between 1980 and 2024. We further filter on papers published in the top-100 venues according to Google Scholar Metrics (accessed 1 October 2025)~\citep{google_scholar_top_publications,google_scholar_metrics}. We exclude papers with more than 20 authors. The resulting sample comprises 4,899,176 papers, 6,501,695 authors, and 65,206 institutions (SI Sections~\ref{sec:top_venues} and \ref{sec:openalex_dataset}).

\paragraph{Institutional Prestige.}
We characterize institutional prestige using the 2025 Times Higher Education (THE) World University Rankings~\citep{the_rankings_2025}. We map THE entries to OpenAlex institution records via algorithmic name-matching, after conditioning on country. Manual review identified only 49 mistakes ($\approx$2.5\%) among the top 2,000 institutions, which were corrected. 

We restrict analysis to the authors' last-known institution recorded in OpenAlex. 
When an author lists multiple current affiliations, we assign the author to the highest-ranked institution in the top-2000 list. This procedure allows us to map 2,975,085 authors ($\approx$46\% of the filtered sample) to a THE-ranked institution (SI Section~\ref{sec:inst_ranking}). The remaining institutions are categorized as \textit{unranked}.

\paragraph{Author Gender.}
Author gender was inferred from first names using a probabilistic first-name classifier implemented in the \texttt{nomquamgender} package~\citep{van2023open}. For each OpenAlex author record we extracted \texttt{display\_name} and any \texttt{display\_name\_alternatives}, extracted first-name tokens, classified each token, and aggregated token-level labels via a majority-vote rule (tie-breaking on the primary display name). Labels were mapped to the categories \textit{men}, \textit{women}, and \textit{unknown}. We also record a \texttt{conflicting} flag for authors whose name variants produced both male and female labels. Automatically inferred labels cover the full author population: 2,750,961 (42\%) men, 1,674,873 (26\%) women, and 2,075,861 (32\%) unknown. To validate this procedure for high-impact cases, we manually reviewed and resolved unknown labels for authors at THE top-25 institutions with more than 100 papers in our top-venue sample (SI Section~\ref{sec:author_gender}).

\paragraph{Collaboration Network.}
We construct an undirected collaboration network from papers: nodes are authors, and an unweighted edge between two authors appears if they co-authored at least one paper in our filtered dataset. For group-level collaborator composition analyses, we attribute each co-author to an institution and THE rank (or Unranked) using the co-author’s last-known institution recorded in OpenAlex.

\paragraph{Metrics.}
We use the following metrics to capture aspects of productivity and collaboration:

\begin{enumerate}
  \item \textbf{CCDF (complementary cumulative distribution function).} 
  For an author-level variable $X$ (number of papers in top venues or number of collaborators), we evaluate the empirical CCDF
  \begin{equation}
  \label{eq:ccdf}
  P(X > x) \;=\; \frac{N_{>x}}{N},
  \end{equation}
  where $N_{>x}$ is the number of authors with $X>x$ and $N$ is the total number of authors    
(Fig~\ref{fig:prestige-THE-CCDF}). 
  The CCDF gives the probability that a randomly chosen author has more than $x$ papers (or collaborators), estimated empirically as the fraction of authors exceeding that threshold.

  \item \textbf{Prestige advantage.} 
    Prestige advantage measures how much more likely it is to find a prolific author at a top-$k$ institution than outside of it. Formally, for a threshold $x$, it is defined as:
  \begin{equation}
  \label{eq:prestige-adv}
  A(x,k) \;=\; \frac{P(X > x \mid \text{Top-}k)}{P(X > x \mid \text{non–Top-}k)}.
  \end{equation}
  Here $P(X>x \mid \cdot)$ denotes the probability that a randomly chosen author in the specified group exceeds the threshold $x$. A value $A(x,k)>1$ indicates that authors exceeding the threshold $x$ are overrepresented at top-$k$ institutions relative to the rest (Fig~\ref{fig:prestige-THE-ranking}). This formulation applies to both productivity and collaboration: $X$ represents either the number of papers or the number of collaborators, with $x$ denoting the corresponding threshold. Fig.~\ref{fig:prestige-gender-THE-ranking} panels (a) and (c) report prestige advantage calculated separately for men and women authors for paper counts and collaboration counts, respectively.

  \item \textbf{Relative gender gap in prestige advantage.} To compare how prestige advantage differs by gender we compute the proportional difference
  \begin{equation}
  \label{eq:gender-delta}
  \Delta(x,k) \;=\; \frac{A_{\text{men}}(x,k) - A_{\text{women}}(x,k)}{A_{\text{women}}(x,k)},
  \end{equation}
  where $A_{g}(x,k)$ is Eq.~\ref{eq:prestige-adv} evaluated for gender $g \in \{\text{men}, \text{women}\}$.  A value of $\Delta(x,k)=0$ indicates parity between genders. Positive $\Delta$ indicates that institutional prestige benefits men more than women, while negative values indicate the opposite. Heat maps visualize $\Delta(x,k)$ across number of papers (or collaborators) thresholds $x$ and rank thresholds $k$ (Fig~\ref{fig:prestige-gender-THE-ranking} panels (b) and (d)).

  \item \textbf{Local clustering coefficient.} For each author $i$ we compute the standard clustering coefficient
  \[
  C_i = \frac{2T_i}{d_i(d_i-1)},
  \]
  where $d_i$ is author $i$’s degree and $T_i$ is the number of triangles containing $i$. The clustering coefficient captures the extent to which an author's collaborators are also connected to one another and ranges from 0 to 1, with higher values indicating a denser  collaboration structure among an author's co-authors. We construct the co-authorship network as an unweighted, undirected graph in which two authors are connected if they have co-authored at least one paper in the dataset. Clustering coefficients are computed using the NetworkX implementation \cite{hagberg2007exploring}. For authors with degree less than two, clustering is defined to be zero.
  We summarize $C_i$ by gender and institutional prestige groups, reporting mean values with 95\% confidence intervals. We estimate clustering in two complementary ways: (A) on the full co-authorship network, and (B) on prestige-restricted subgraphs induced by authors affiliated with institutions ranked $\le k$ (Fig~\ref{fig:cluesting_gender_prestige}).

  \item \textbf{Collaborator prestige.} For each focal group (gender $\times$ institutional prestige class) we partition collaborators into four mutually exclusive categories: collaborators at more prestigious institutions, collaborators in the same rank class but at a different institution, collaborators at the same institution (internal), and collaborators at less prestigious institutions. Shares are computed from aggregated collaborator counts (not per-author averages), ensuring that results reflect the full volume of collaborative ties. Results are displayed as stacked bars (Fig~\ref{fig:Collaboration_prestige_mix}).
\end{enumerate}



\paragraph{Data limitations and robustness checks.}
Our analysis relies on several operational choices. First, we use authors’ last-known institutional affiliation as a proxy for institutional position, which may not fully capture career mobility over time. Second, linking institutional records to THE rankings requires name matching, which may introduce minor inconsistencies despite manual verification of ambiguous cases. Third, gender is inferred from first names using a scalable approach commonly employed in large bibliometric studies; authors with ambiguous or unknown labels are retained in the dataset but excluded from analyses requiring gender classification. 

We assess the robustness of our results to these design choices through a series of sensitivity analyses reported in the Supplementary Information. 

}


\showmatmethods{} 

\acknow{This work was supported by a Keston Exploratory Research Award awarded by the USC Information Sciences Institute (ISI). We thank the ISI Center on Knowledge Graphs for providing the computational resources used in this study.}

\showacknow{} 


\bibliography{references}

\newpage
\clearpage


\setcounter{section}{0}
\renewcommand{\thesection}{\arabic{section}}
\setcounter{subsection}{0}
\renewcommand{\thesubsection}{\thesection.\alph{subsection}}

\renewcommand\thefigure{S\arabic{figure}}
\setcounter{figure}{0}

\onecolumn

\section*{Supplementary Information}
\section{Identification of high-impact publication venues}
\label{sec:top_venues}

As a 
proxy of scholarly impact, we consider publication in one of the Google Scholar top-100 venues (accessed on October 1, 2025). Google Scholar ranks venues according to their five-year h-index (h5-index),\footnote{Top Publications list: \url{https://scholar.google.com/citations?view_op=top_venues}}
defined as the largest number $h$ such that $h$ articles published in a venue during this period each received at least $h$ citations.\footnote{\url{https://scholar.google.com/intl/en/scholar/metrics.html}}
%
This set of venues includes journals and proceedings that span a wide range of disciplines, including interdisciplinary research (e.g., \textit{Nature}, \textit{Science}), biomedical sciences (e.g., \textit{The Lancet}, \textit{Cell}), physical sciences (e.g., \textit{Journal of the American Chemical Society}, \textit{Chemical Reviews}), environmental science (e.g., \textit{Energy \& Environmental Science}, \textit{Applied Energy}), and computer science and engineering (e.g., NeurIPS, ICML, ICLR). 
These publications employ highly selective editorial and  peer review processes, ensuring that only research that meets rigorous standards of quality and impact is published (Table~\ref{tab:top100_gs_venues}). 

We manually matched each of the 100 publication venues to its corresponding OpenAlex source identifier. All subsequent analyses are restricted to papers published in these venues.

\begin{figure*}[h]
\centering
\begin{tabular}{cc}
    \includegraphics[width=0.48\linewidth]{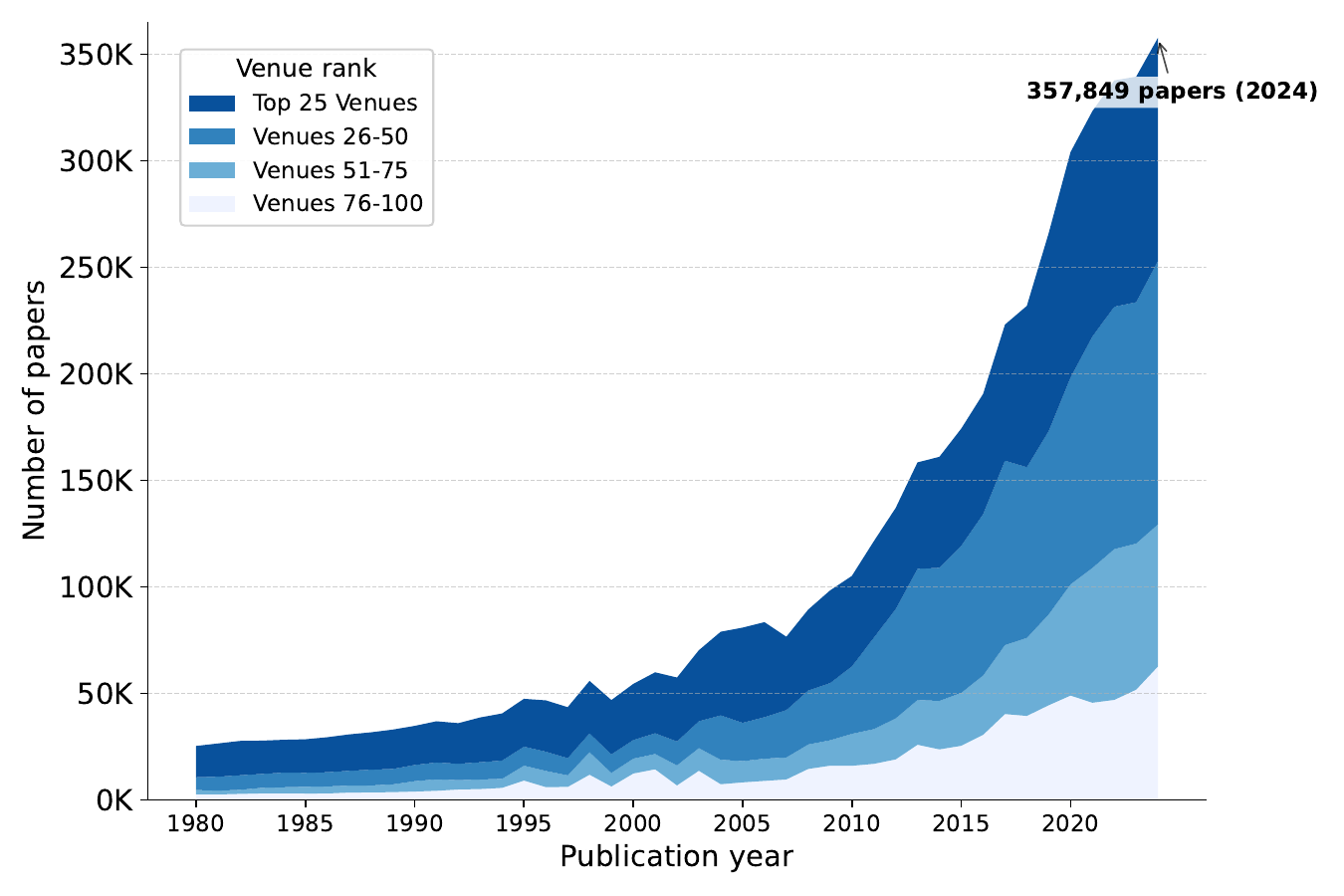} &
    \includegraphics[width=0.48\linewidth]{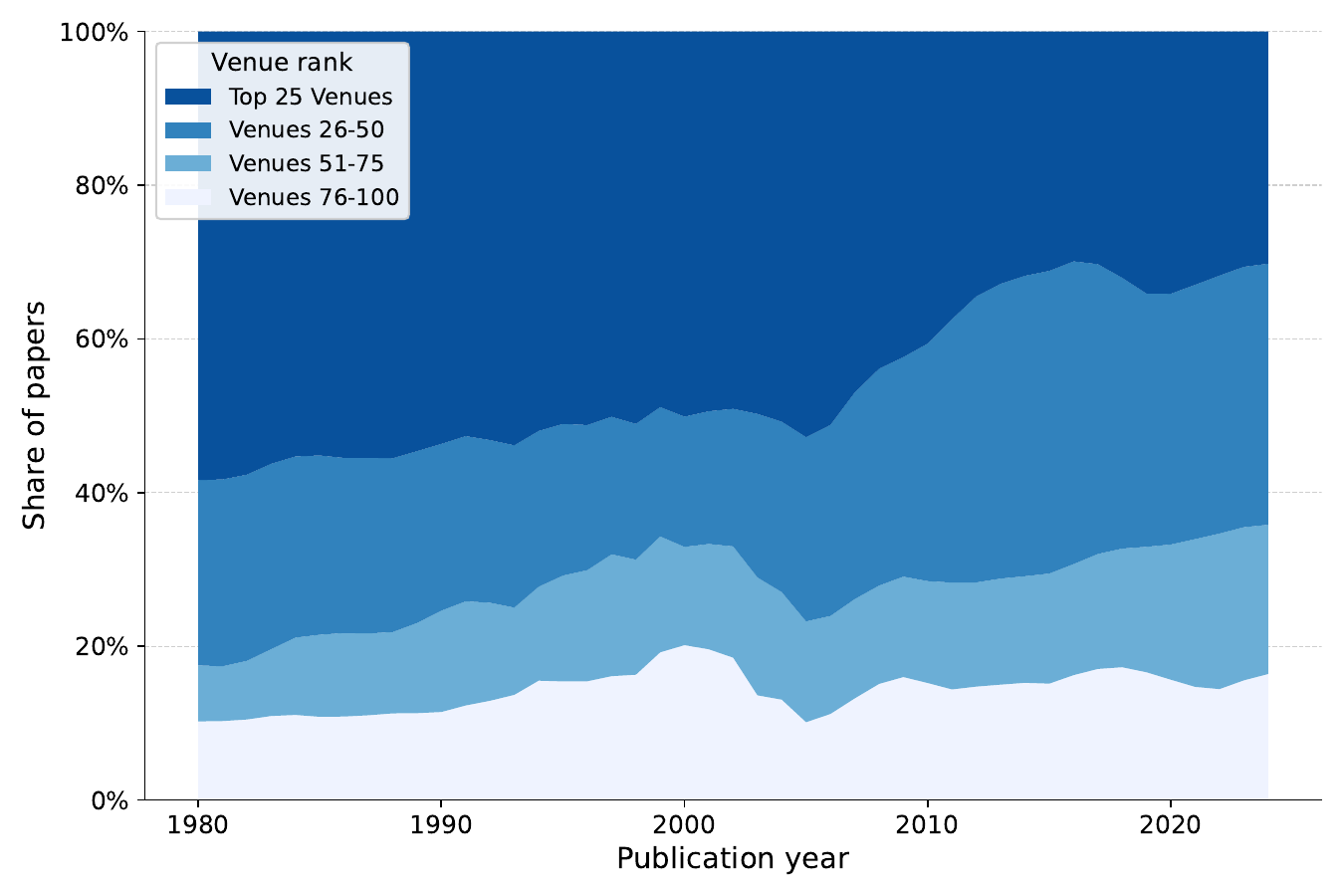} \\
    (a) & (b)
\end{tabular}
\caption{\textbf{Publication trends of the top-100 Google Scholar venues (1980--2024).}
(a) Number of papers published annually in the Google Scholar top-100 venues, grouped by venue rank. 
This growth in the number of papers reflects both an increase in publication output and the fact that the set of venues is defined based on the October 2025 ranking and held fixed across time; some venues in this set did not exist in earlier years and therefore contribute only to later counts. 
(b) Annual share of papers by venue rank group. While top-25 venues account for nearly 60\% of publications in the early 1980s, their share declines to approximately 30\% by 2024, with publication volume becoming more evenly distributed across venue ranks, though not uniformly so.}
\label{fig:dist_works}
\end{figure*}

\paragraph{Temporal consistency of venue selection.}
The set of top 100 venues is defined using the Google Scholar Metrics ranking as of October 1, 2025 and is held fixed across the entire observation window (1980–2024). As a result, some venues included in this set were not active in earlier years and therefore contribute only to later publication counts. Consequently, the number of eligible venues, and thus the volume of publications, increases over time (Fig.~\ref{fig:dist_works}). 
This design provides a consistent, externally defined benchmark for high-impact publication outlets, while introducing temporal variation in venue availability. Although publication practices and standards may evolve over time, fixing the venue set allows for a stable point of comparison across periods, rather than attempting to reconstruct historically varying notions of “top” venues.


\begin{table*}[p]
\centering
\scriptsize

\caption{\textbf{Top 100 Google Scholar venues used to define the filtered publication sample.}
Venues are listed alphabetically within each column for compact display. Only OpenAlex works published in these venues were retained in the analysis.}
\label{tab:top100_gs_venues}

\begin{minipage}[t]{0.24\textwidth}
\raggedright
\begin{itemize}\setlength\itemsep{1pt}
\item AAAI Conference on Artificial Intelligence
\item ACM Computing Surveys
\item ACS Applied Materials \& Interfaces
\item ACS Catalysis
\item ACS Energy Letters
\item ACS Nano
\item Advanced Energy Materials
\item Advanced Functional Materials
\item Advanced Materials
\item Advanced Science
\item Angewandte Chemie International Edition
\item Applied Catalysis B: Environmental
\item Applied Energy
\item Applied Sciences
\item Bioresource Technology
\item BMJ
\item Cell
\item Cells
\item Chemical Engineering Journal
\item Chemical Reviews
\item Chemical Society Reviews
\item Chemosphere
\item Circulation
\item Clinical Infectious Diseases
\item Conference on Empirical Methods in Natural Language Processing (EMNLP)
\end{itemize}
\end{minipage}
\hfill
\begin{minipage}[t]{0.24\textwidth}
\raggedright
\begin{itemize}\setlength\itemsep{1pt}
\item Construction and Building Materials
\item Coordination Chemistry Reviews
\item Energy
\item Energy \& Environmental Science
\item Energy Storage Materials
\item Environmental Pollution
\item Environmental Science and Pollution Research
\item Environmental Science \& Technology
\item European Conference on Computer Vision
\item European Heart Journal
\item Expert Systems with Applications
\item Frontiers in Immunology
\item Frontiers in Psychology
\item Gastroenterology
\item IEEE Access
\item IEEE Internet of Things Journal
\item IEEE Transactions on Industrial Informatics
\item IEEE Transactions on Pattern Analysis and Machine Intelligence
\item IEEE/CVF Conference on Computer Vision and Pattern Recognition
\item IEEE/CVF International Conference on Computer Vision
\item International Conference on Learning Representations
\item International Conference on Machine Learning
\item International Journal of Biological Macromolecules
\item International Journal of Environmental Research and Public Health
\item International Journal of Hydrogen Energy
\end{itemize}
\end{minipage}
\hfill
\begin{minipage}[t]{0.24\textwidth}
\raggedright
\begin{itemize}\setlength\itemsep{1pt}
\item International Journal of Molecular Sciences
\item JAMA
\item JAMA Network Open
\item Joule
\item Journal of Business Research
\item Journal of Cleaner Production
\item Journal of Clinical Oncology
\item Journal of Environmental Management
\item Journal of Hazardous Materials
\item Journal of Materials Chemistry A
\item Journal of Medical Internet Research
\item Journal of Retailing and Consumer Services
\item Journal of the American Chemical Society
\item Journal of the American College of Cardiology
\item Meeting of the Association for Computational Linguistics (ACL)
\item Molecules
\item Morbidity and Mortality Weekly Report
\item Nano Energy
\item Nature
\item Nature Biotechnology
\item Nature Communications
\item Nature Energy
\item Nature Genetics
\item Nature Materials
\item Nature Medicine
\end{itemize}
\end{minipage}
\hfill
\begin{minipage}[t]{0.24\textwidth}
\raggedright
\begin{itemize}\setlength\itemsep{1pt}
\item Nature Nanotechnology
\item Neural Information Processing Systems
\item Nucleic Acids Research
\item Nutrients
\item Physical Review Letters
\item PLOS ONE
\item Proceedings of the National Academy of Sciences
\item Renewable and Sustainable Energy Reviews
\item Renewable Energy
\item Science
\item Science Advances
\item Science of The Total Environment
\item Scientific Reports
\item Sensors
\item Signal Transduction and Targeted Therapy
\item Small
\item Sustainability
\item Technological Forecasting and Social Change
\item The Astrophysical Journal
\item The Lancet
\item The Lancet Infectious Diseases
\item The Lancet Oncology
\item The New England Journal of Medicine
\item Trends in Food Science \& Technology
\item Water Research
\end{itemize}
\end{minipage}

\end{table*}

\section{OpenAlex Data Preprocessing}
\label{sec:openalex_dataset}

We draw bibliometric data from OpenAlex, a public catalog of the global scholarly research system. OpenAlex aggregates and standardizes metadata on scholarly papers, authors, venues, institutions, and citations from a wide range of underlying sources, including Crossref, PubMed, PubMed Central, ORCID, ROR, Unpaywall, and subject-area repositories such as arXiv. Its open data model, broad disciplinary coverage, and transparent update process make it well suited for large-scale analyses of scientific production.

We use the OpenAlex data snapshot released on July 21, 2025, accessed via the bulk data dump. From the full corpus, we apply the following filters. First, we restrict our sample to papers published in the set of 100 high-impact publication venues (see Section~\ref{sec:top_venues}). Second, we limit the publication year to the period 1980–-2024. 
Third, we exclude papers with more than 20 authors to reduce the influence of large-scale consortia 
which may distort analyses of standard collaboration practices.

After filtering, the resulting dataset includes 4,899,176 unique papers, associated with 6,501,695 unique authors and 65,206 unique institutions. 

\section{Institutional Ranking and Author Affiliation}
\label{sec:inst_ranking}


To characterize the institutional prestige of author's affiliation, we use the Times Higher Education (THE) World University Rankings 2025~\cite{the_rankings_2025}. The THE rankings provide a global comparison of research-intensive universities across multiple dimensions, including teaching, research environment, research quality, international outlook, and industry engagement. Rankings are derived from 18 performance indicators and are based on institutional data reported by universities and standardized across indicators using distribution-based methods. We use the 2025 rankings as published on September 23, 2024, and last updated on October 1, 2025.

We restrict attention to the top 2,000 institutions listed in the 2025 THE World University Rankings. To link ranked institutions to the OpenAlex database, we first obtained a downloadable version of the THE rankings containing institution names and ranks. We then performed an algorithmic name-matching procedure to align THE institutions with OpenAlex institution records, conditioning on institutional country to reduce ambiguity. All matches were manually reviewed, and mismatches were corrected, including cases where a single OpenAlex institution identifier was erroneously assigned to multiple universities or where a matched identifier yielded implausibly few or no publications. In total, 49 institution identifiers were corrected, corresponding to approximately 2.5\% of the top 2,000 ranked institutions.

Author–institution assignments are based on the ``last known institution'' field provided in the OpenAlex data dump. When an author is associated with multiple current institutions, we assign the author to the highest-ranked institution among those appearing in the top 2,000 THE rankings, if applicable. Authors whose listed institutions do not appear in the top 2,000 are treated as unranked for the purposes of institutional rank assignment.

Applying this procedure, 2,975,085 authors are successfully mapped to a ranked institution, representing approximately 46\% of the 6,501,695 authors in the filtered OpenAlex sample. Institutional rank is treated as a fixed external attribute and is used to characterize differences in publication and collaboration patterns across the global hierarchy of research institutions.

\section{Author gender inference}
\label{sec:author_gender}


Author gender was inferred using a first name–based classification procedure applied to OpenAlex author metadata. Our starting point was the author dataset described above, containing 6,501,695 unique authors with OpenAlex author identifiers (\texttt{author\_id}) and associated bibliometric information.

To obtain author name information, we matched each \texttt{author\_id} to the corresponding record in the OpenAlex authors snapshot released on July 21, 2025. From these records we extracted the fields \texttt{display\_name} and \texttt{display\_name\_alternatives}. The \texttt{display\_name} field contains the canonical author name provided by OpenAlex, while \texttt{display\_name\_alternatives} includes alternative name formats observed across papers (for example, variations in name ordering or abbreviated initials).

For each author, we extracted first names from both the primary display name and all available alternative name variants. When names appeared in the format “Last, First”, the first-name component was extracted accordingly. Only the first whitespace-delimited token of each name was used for gender inference. This procedure generated a set of candidate first names for each author.

Gender classification for first names was performed using the \texttt{nomquamgender} package~\cite{van2023open}, which implements a probabilistic first-name gender classifier trained on large-scale name–gender associations. The model returns one of three labels for each first name: male, female, or unknown. For consistency with the vocabulary used in this study, these labels were mapped to the categories \textit{men}, \textit{women}, and \textit{unknown}. First-name–based gender inference is widely used in large-scale bibliometric studies where self-reported demographic data are unavailable, and provides a practical approach for characterizing aggregate gender patterns in scholarly populations.

For each author, gender labels inferred from all available first names were aggregated using a majority-vote rule. Let $g_m$ denote the number of first names classified as male and $g_w$ the number classified as female, ignoring names labeled unknown. Gender assignment proceeded as follows:

\begin{enumerate}
\item If $g_m > g_w$, the author was assigned \textit{men}.
\item If $g_w > g_m$, the author was assigned \textit{women}.
\item If $g_m = g_w$ and both categories were present, the classification of the primary \texttt{display\_name} was used as a tie-breaker when available.
\item If the tie could not be resolved, or if all name classifications were \textit{unknown}, the author was assigned \textit{unknown}.
\end{enumerate}

In addition to the inferred gender label, we recorded a binary indicator (\texttt{conflicting}) identifying authors for whom both male and female classifications were observed among their name variants. This flag allows ambiguous cases to be examined separately in downstream analyses. We manually inspected a random sample of 100 authors flagged as conflicting and found that the inferred gender label was consistent with the available name evidence in all cases.

Applying this procedure to the full author dataset produced the following distribution of inferred gender labels: 2,750,961 authors (42\%) were classified as men, 1,674,873 authors (26\%) as women, and 2,075,861 authors (32\%) remained classified as unknown. Authors labeled as unknown were retained in the dataset but excluded from analyses requiring binary gender classification.

\paragraph{Manual validation for high-impact authors.}
To validate gender inference and to increase coverage of highly productive authors at elite institutions, we conducted a targeted manual review of all authors with \textit{unknown} gender who were affiliated with the top 25 institutions according to the Times Higher Education World University Rankings and had more than 100 papers in our dataset. This subset comprised 238 authors.

For each author, we located their OpenAlex author page using the \texttt{author\_id}, then performed manual verification by searching for the author using their display name and institutional affiliation. In cases where names were ambiguous, searches were refined using titles of highly cited papers. Gender was inferred using publicly available information such as pronouns listed on institutional faculty pages or in reputable news articles, as well as profile photographs on faculty pages, Google Scholar profiles, or associated ORCID records.

Among authors in the top 25 institutions with more than 100 papers and known gender labels, the gender distribution was 85\% men (758) and 15\% women (131). Manual labeling of the subset of 238 previously unknown authors yielded a similar distribution: 84\% men (201) and 16\% women (37). 
These results indicate that gender inference procedure does not materially distort observed gender balance in the data.

The distribution of authors across institutional rank classes and gender is summarized in Table~\ref{tab:author_composition_by_rank}. The distribution of authors by productivity and collaboration thresholds is reported in Table~\ref{tab:authors_count_by_works} and Table~\ref{tab:authors_count_by_collabs}, respectively.

\section{Dataset Descriptive Statistics}
\label{sec:data_stats}

\begin{figure*}[ht]
    \centering
    \includegraphics[width=0.48\linewidth]{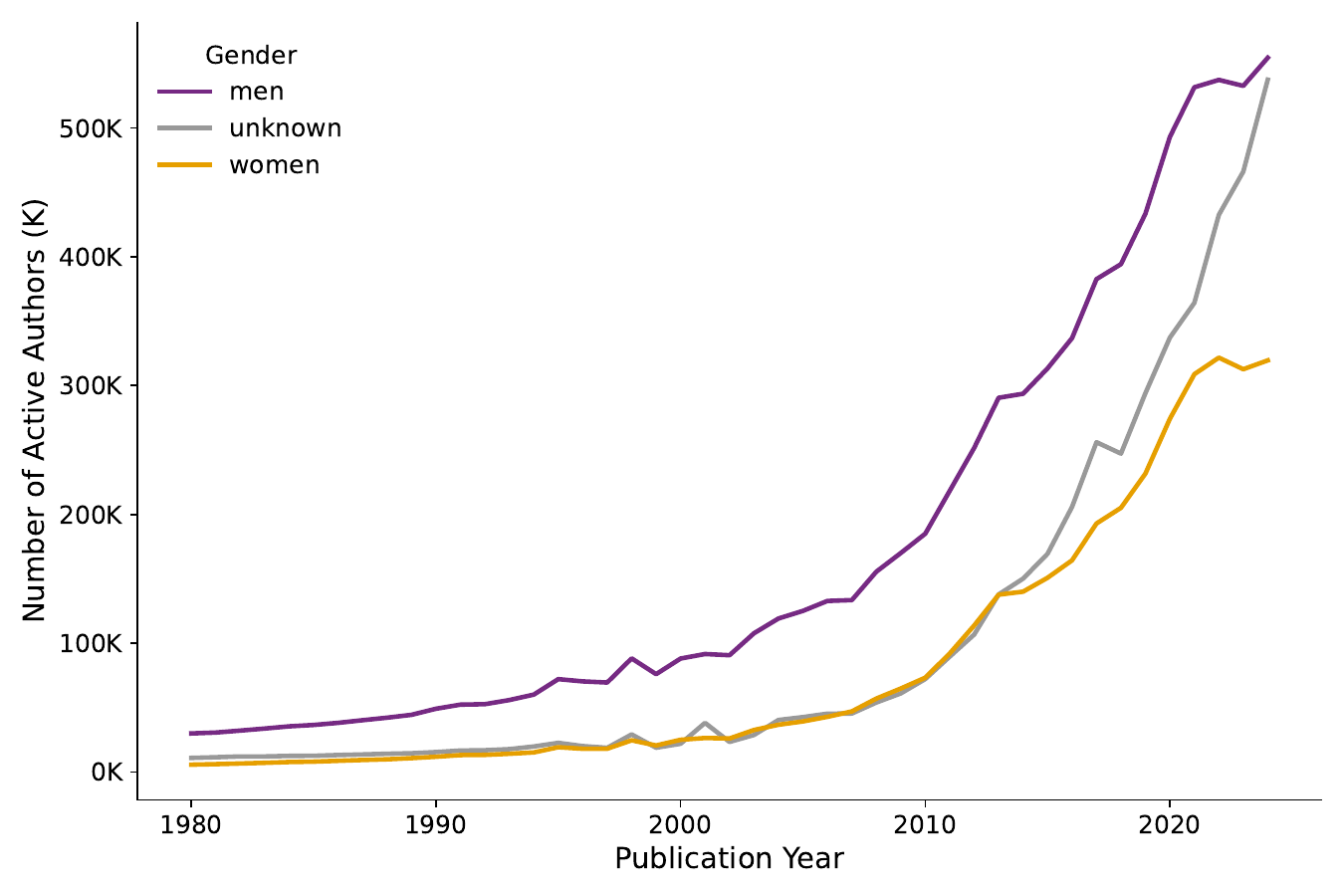}
    \includegraphics[width=0.48\linewidth]{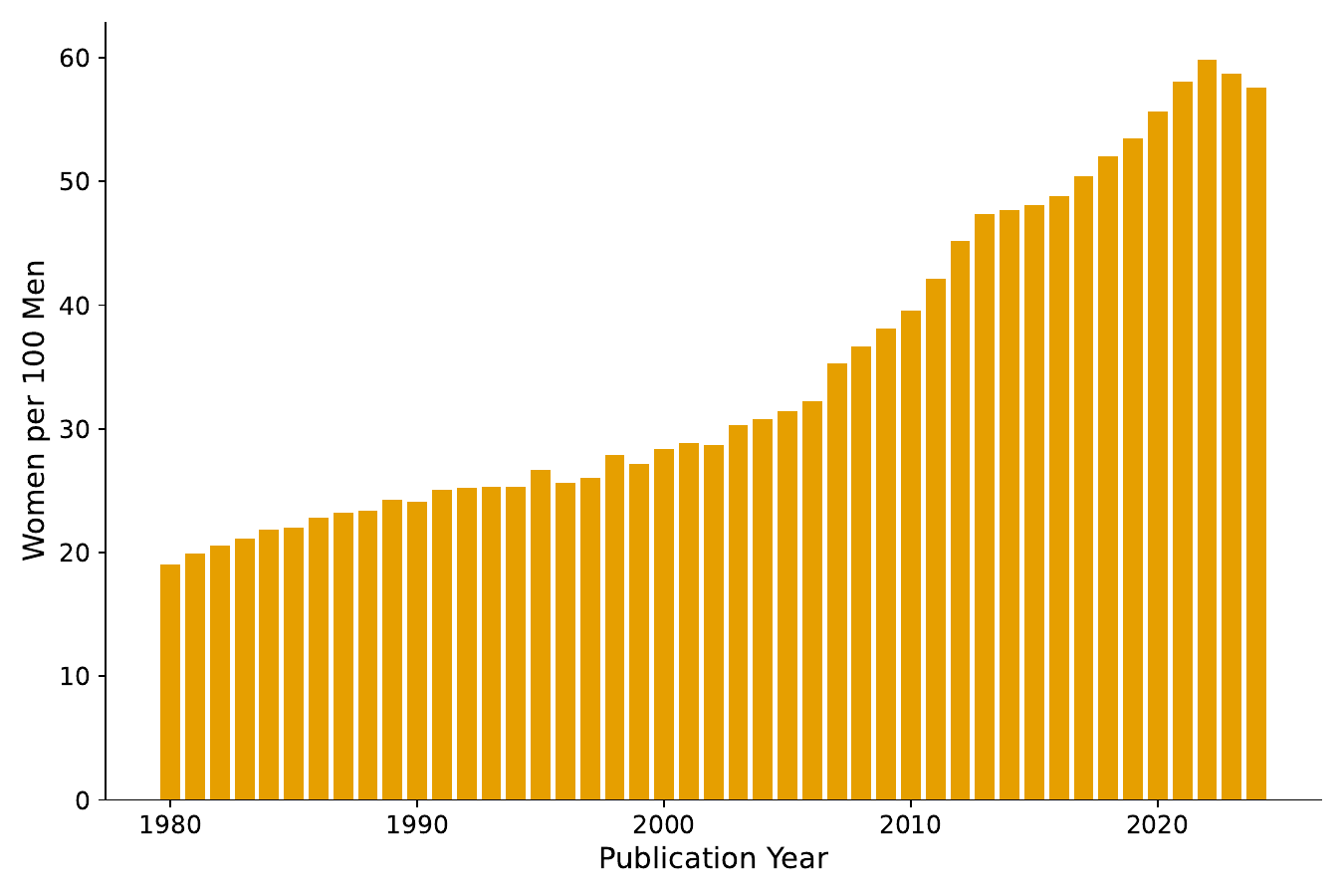}
    \caption{\textbf{Growth in the representation of women authors.}
    (a) Number of active authors by gender over time in the top-100 venues. (b) Women per 100 men. Although women remain underrepresented, their representation rises from about 20 per 100 men in 1980 to nearly 60 by 2024. 
    }
    \label{fig:women_by_year}
\end{figure*}

Figure~\ref{fig:women_by_year} summarizes the temporal evolution of gender representation in the dataset. Panel (a) shows the number of active authors by gender, defined as authors publishing in a given year in the top-100 venues, revealing substantial growth for both men and women over time. Panel (b) reports the ratio of women authors per 100 men, indicating a steady increase in women’s representation---from fewer than 20 per 100 men in 1980 to nearly 60 by 2024. Although women remain underrepresented overall, this trend reflects a gradual narrowing of the gender gap. Authors with unknown gender follow similar patterns in earlier years but diverge in later periods, likely due to changes in data coverage and limitations of name-based gender inference.

\begin{figure*}[t]
\centering
\setlength{\abovecaptionskip}{4pt}
\setlength{\belowcaptionskip}{0pt}

\begin{minipage}[t]{0.46\textwidth}
    \vspace{0pt}
    \centering
    \includegraphics[width=\linewidth]{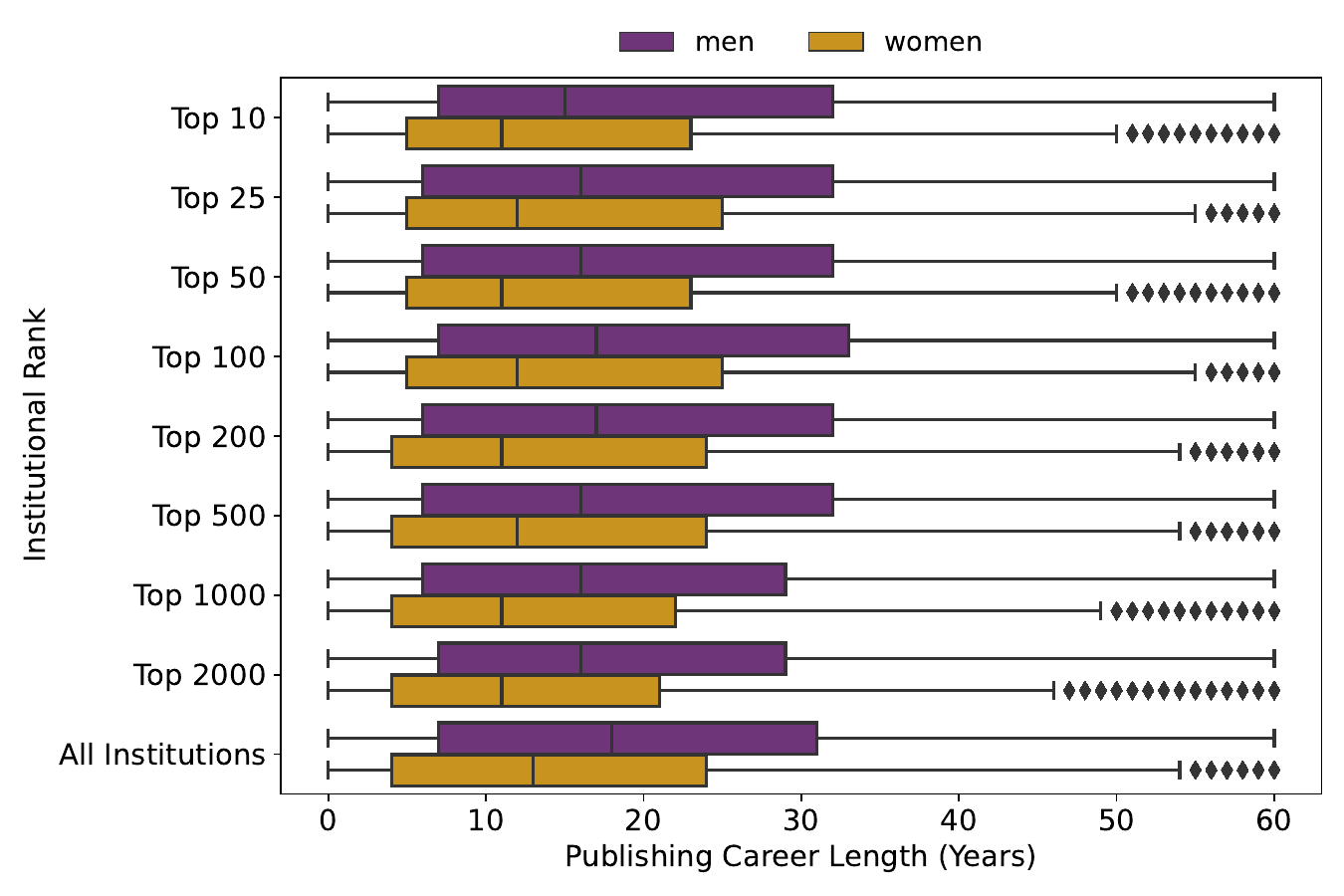}
\end{minipage}
\hfill
\begin{minipage}[t]{0.52\textwidth}
    \vspace{0pt}
    \centering
    \scriptsize
    \setlength{\tabcolsep}{3.5pt}
    \renewcommand{\arraystretch}{1.15}

    \begin{tabular}{lcc cc}
        \toprule
        & \multicolumn{2}{c}{\textbf{Author Count}} & \multicolumn{2}{c}{\textbf{Median Works Count}} \\
        \cmidrule(lr){2-3} \cmidrule(lr){4-5}
        \textbf{Institutional Rank} & \textbf{Men} & \textbf{Women} & \textbf{Men} & \textbf{Women} \\
        \midrule
        Top 10            & 52,664    & 33,805    & 15 & 11 \\
        Top 25            & 67,415    & 45,260    & 16 & 12 \\
        Top 50            & 78,407    & 49,331    & 16 & 11 \\
        Top 100           & 120,258   & 78,435    & 17 & 12 \\
        Top 200           & 179,990   & 113,893   & 17 & 11 \\
        Top 500           & 292,081   & 190,146   & 16 & 12 \\
        Top 1000          & 223,066   & 146,612   & 16 & 11 \\
        Top 2000          & 226,429   & 143,778   & 16 & 11 \\
        All Institutions  & 1,339,614 & 820,735   & 18 & 13 \\
        \bottomrule
    \end{tabular}
\end{minipage}

\caption{\textbf{Career length and author composition by institutional rank.}
The left panel shows the distribution of author career length (tenure), defined as the number of years between an author’s first and last publication in the full OpenAlex dataset. Boxes represent the interquartile range (25th–75th percentiles) with the median indicated; whiskers extend to 1.5$\times$IQR, and points beyond are shown as outliers. To mitigate data artifacts, authors with career lengths exceeding 60 years are excluded. The right panel reports author counts and median number of papers by institutional rank and gender.}
\label{fig:career_length_and_table}
\end{figure*}

Figure~\ref{fig:career_length_and_table} shows the distribution of author career length, defined as the number of years between an author’s first and last recorded publication in the full OpenAlex dataset (not restricted to publications in the top-100 venues), providing a more complete measure of scientific tenure. We exclude authors with careers longer than 60 years to mitigate implausibly long career durations arising from errors in publication year metadata. This threshold removes a small fraction of the data (6.2\% of men and 3.2\% of women) while preserving the overall structure of the distribution. Career length distributions are summarized using box plots, where the interquartile range (25th–75th percentiles) captures central variation, the median indicates the central tendency, and outliers reflect highly atypical publication histories.

While women tend to have shorter careers than men, consistent with prior work showing higher attrition among women in academia~\cite{huang2020historical}, we do not observe systematic differences across institutional ranks within each gender group. Because our analysis compares authors within gender, these differences in overall career length are unlikely to explain the observed patterns in prestige advantage. These patterns suggest that differences in prestige advantage are unlikely to be driven by systematic variation in career duration.

Figure~\ref{fig:dist_works} summarizes the temporal evolution of publications in the fixed set of top-100 venues. Panel (a) shows a substantial increase in annual output, from approximately 25,000 papers in the early 1980s to over 350,000 by 2024. Panel (b) illustrates the distribution of papers across venue tiers. 
Because the venue set is defined using the October 2025 Google Scholar ranking and held fixed over time, some venues were not active in earlier years and contribute only to later counts. As a result, the observed growth reflects both increasing publication output and changes in venue availability.

\begin{table}[ht]
\centering
\small
\caption{Author composition by institutional rank group and gender in the filtered top-100 venue sample (1980--2024). Gender representation percentages are computed relative to the total number of authors within each rank class. The final column reports the number of women per 100 men.}
\label{tab:author_composition_by_rank}
\begin{tabular}{lccccc}
\toprule
\textbf{Institutional Rank} & \textbf{\% Authors} & \textbf{\% Men} & \textbf{\% Women} & \textbf{\% Unknown} & \textbf{Women per 100 Men} \\
\midrule
Top 10            & 2\%   & 50\% & 30\% & 20\% & 61 \\
Top 25            & 5\%   & 45\% & 29\% & 26\% & 63 \\
Top 50            & 8\%   & 45\% & 28\% & 28\% & 62 \\
Top 100           & 12\%  & 44\% & 28\% & 28\% & 62 \\
Top 200           & 19\%  & 45\% & 28\% & 27\% & 62 \\
Top 500           & 30\%  & 44\% & 28\% & 28\% & 62 \\
Top 1000          & 38\%  & 44\% & 27\% & 29\% & 63 \\
Top 2000          & 46\%  & 44\% & 28\% & 28\% & 63 \\
All Institutions  & 100\% & 42\% & 26\% & 32\% & 61 \\
\bottomrule
\end{tabular}
\end{table}

\begin{table*}[ht]
\centering
\scriptsize
\caption{\boldmath Number of authors with more than $x$ papers in our filtered top-100 venue sample (1980-2024), by institutional rank group and gender. Numbers derived from the OpenAlex bulk snapshot (21 July 2025), filtered to papers appearing in the top-100 Google Scholar venues (accessed 1 Oct 2025) and publication years 1980-2024; papers with $>20$ authors were excluded. Counts are cumulative (not discrete).}
\label{tab:authors_count_by_works}

\resizebox{\textwidth}{!}{%
\begin{tabular}{@{} l *{12}{S[table-format=7]} @{}}
\toprule
 & {${>}$0} & {${>}$1} & {${>}$2} & {${>}$5} & {${>}$10} & {${>}$25} & {${>}$50} & {${>}$75} & {${>}$100} & {${>}$125} & ${>}$150 & ${>}$200 \\
\midrule

\multicolumn{13}{@{}l}{\textbf{All Institutions}} \\
\quad All authors   & 6501695 & 3194950 & 2105982 & 974905 & 446019 & 118219 & 34692 & 15360 & 8318 & 4947 & 3230 & 1595 \\
\quad Men           & 2751089 & 1477729 & 1019092 & 506378 & 246826 & 71756 & 22554 & 10482 & 5854 & 3560 & 2352 & 1186 \\
\quad Women         & 1674881 & 777211  & 481973  & 193235 & 76271  & 16372 & 4247  & 1780  & 955  & 570  & 369  & 197  \\

\midrule
\multicolumn{13}{@{}l}{\textbf{Top 10}} \\
\quad All authors   & 115657 & 68119 & 47712 & 23960 & 12488 & 4556 & 1773 & 942 & 558 & 344 & 237 & 117 \\
\quad Men           & 57824  & 37186 & 27358 & 15000 & 8436  & 3383 & 1395 & 770 & 474 & 293 & 201 & 103 \\
\quad Women         & 35166  & 19044 & 12367 & 5331  & 2437  & 753  & 274  & 140 & 84  & 51  & 36  & 14  \\

\midrule
\multicolumn{13}{@{}l}{\textbf{Top 25}} \\
\quad All authors   & 289369 & 166108 & 115807 & 58231 & 30071 & 10447 & 3870 & 1947 & 1127 & 703 & 457 & 225 \\
\quad Men           & 130965 & 82140  & 60202  & 32970 & 18370 & 7107  & 2866 & 1561 & 959  & 603 & 394 & 195 \\
\quad Women         & 82337  & 43493  & 28219  & 12353 & 5675  & 1725  & 595  & 281  & 168  & 100 & 63  & 30  \\

\midrule
\multicolumn{13}{@{}l}{\textbf{Top 50}} \\
\quad All authors   & 484807 & 276193 & 192158 & 96120 & 49204 & 16429 & 5878 & 2914 & 1675 & 1049 & 685 & 348 \\
\quad Men           & 215575 & 133875 & 97847  & 53389 & 29552 & 11007 & 4255 & 2248 & 1360 & 862 & 561 & 291 \\
\quad Women         & 133519 & 69702  & 45120  & 19396 & 8748  & 2490  & 841  & 377  & 219  & 135 & 84  & 39  \\

\midrule
\multicolumn{13}{@{}l}{\textbf{Top 100}} \\
\quad All authors   & 778464 & 441831 & 306926 & 153717 & 77801 & 25041 & 8556 & 4146 & 2326 & 1417 & 933 & 477 \\
\quad Men           & 345585 & 213455 & 155892 & 85227  & 46609 & 16535 & 6042 & 3077 & 1802 & 1116 & 735 & 384 \\
\quad Women         & 215228 & 111946 & 72272  & 30947  & 13512 & 3621  & 1136 & 510  & 279  & 169  & 104 & 52  \\

\midrule
\multicolumn{13}{@{}l}{\textbf{Top 200}} \\
\quad All authors   & 1202103 & 675474 & 467083 & 232718 & 116534 & 36427 & 12023 & 5647 & 3095 & 1873 & 1219 & 601 \\
\quad Men           & 539385  & 330206 & 240135 & 130562 & 70545  & 24232 & 8477 & 4155 & 2377 & 1467 & 963 & 484 \\
\quad Women         & 333355  & 171332 & 109944 & 46646  & 19942  & 5053  & 1485 & 641  & 332  & 195  & 118 & 61  \\

\midrule
\multicolumn{13}{@{}l}{\textbf{Top 1000}} \\
\quad All authors   & 2482475 & 1361666 & 933467 & 457533 & 221298 & 63022 & 19221 & 8595 & 4591 & 2712 & 1723 & 835 \\
\quad Men           & 1084709 & 645592  & 464128 & 246334 & 127779 & 39790 & 12888 & 6042 & 3347 & 2027 & 1299 & 651 \\
\quad Women         & 679920  & 341488  & 218668 & 92178  & 37966  & 8480  & 2184 & 903  & 464  & 271  & 167 & 84  \\

\midrule
\multicolumn{13}{@{}l}{\textbf{Top 2000}} \\
\quad All authors   & 2975085 & 1598602 & 1085974 & 523009 & 247965 & 68474 & 20427 & 9020 & 4792 & 2816 & 1787 & 863 \\
\quad Men           & 1321505 & 770228  & 548379  & 285227 & 144590 & 43442 & 13751 & 6367 & 3503 & 2108 & 1348 & 675 \\
\quad Women         & 826410  & 406447  & 257666  & 106461 & 42828  & 9173  & 2292 & 934  & 476  & 275  & 168 & 84 \\

\bottomrule
\end{tabular}%
}
\end{table*}


\begin{table*}[ht]
\centering
\scriptsize
\caption{\boldmath Number of authors with more than $x$ collaborators in our filtered top-100 venue sample (1980--2024), by institutional rank group and gender. Counts are cumulative (not discrete).}
\label{tab:authors_count_by_collabs}

\resizebox{\textwidth}{!}{%
\begin{tabular}{@{} l *{12}{S[table-format=7]} @{}}
\toprule
 & {All authors} & {${>}$0} & {${>}$2} & {${>}$5} & {${>}$10} & {${>}$25} & {${>}$50} & {${>}$75} & {${>}$100} & {${>}$150} & {${>}$200} & {${>}$500} \\
\midrule

\multicolumn{13}{@{}l}{\textbf{All Institutions}} \\
\quad All authors & 6501695 & 6372380 & 5811504 & 4438339 & 2771013 & 1085385 & 441339 & 235097 & 143309 & 66045 & 36237 & 3965 \\
\quad Men         & 2751089 & 2699042 & 2439853 & 1871363 & 1223130 & 523419  & 228618 & 127479 & 80544  & 39137 & 22400 & 2779 \\
\quad Women       & 1674881 & 1654412 & 1526581 & 1184438 & 726468  & 245439  & 84252  & 40141  & 22626  & 9415  & 4837  & 468 \\

\midrule
\multicolumn{13}{@{}l}{\textbf{Top 10}} \\
\quad All authors & 115657 & 113932 & 105098 & 86098 & 61546 & 28015 & 12857 & 7773 & 5242 & 2882 & 1821 & 250 \\
\quad Men         & 57824  & 56850  & 52237  & 43282 & 32039 & 16173 & 8151  & 5192 & 3650 & 2125 & 1385 & 206 \\
\quad Women       & 35166  & 34809  & 32594  & 26699 & 18573 & 7316  & 2907  & 1622 & 1029 & 505  & 311  & 41 \\

\midrule
\multicolumn{13}{@{}l}{\textbf{Top 25}} \\
\quad All authors & 289369 & 285669 & 264090 & 215376 & 150965 & 67835 & 31169 & 18666 & 12505 & 6670 & 4019 & 556 \\
\quad Men         & 130965 & 128924 & 118644 & 97780  & 71323  & 35557 & 17835 & 11288 & 7893  & 4514 & 2843 & 455 \\
\quad Women       & 82337  & 81567  & 76194  & 62416  & 42603  & 16569 & 6683  & 3731  & 2387  & 1167 & 663  & 90 \\

\midrule
\multicolumn{13}{@{}l}{\textbf{Top 50}} \\
\quad All authors & 484807 & 479565 & 443477 & 359340 & 249440 & 111438 & 50790 & 29986 & 19756 & 10324 & 6119 & 816 \\
\quad Men         & 215575 & 212669 & 195505 & 159797 & 115480 & 57200  & 28565 & 17848 & 12303 & 6833  & 4230 & 651 \\
\quad Women       & 133519 & 132458 & 123776 & 100730 & 67859  & 25987  & 10247 & 5548  & 3461  & 1660  & 924  & 113 \\

\midrule
\multicolumn{13}{@{}l}{\textbf{Top 100}} \\
\quad All authors & 778464 & 770637 & 712182 & 574017 & 395121 & 175785 & 79503 & 46278 & 30091 & 15337 & 8978 & 1115 \\
\quad Men         & 345585 & 341250 & 313416 & 254810 & 182809 & 90125  & 44635 & 27397 & 18520 & 9945  & 6054 & 854 \\
\quad Women       & 215228 & 213666 & 199513 & 161384 & 107334 & 40855  & 15723 & 8291  & 5047  & 2353  & 1282 & 140 \\

\midrule
\multicolumn{13}{@{}l}{\textbf{Top 200}} \\
\quad All authors & 1202103 & 1190627 & 1096747 & 873623 & 593863 & 261178 & 116554 & 66842 & 42876 & 21372 & 12243 & 1436 \\
\quad Men         & 539385  & 533015  & 487916  & 392352 & 278226 & 135839 & 66349  & 40011 & 26639 & 13939 & 8253  & 1091 \\
\quad Women       & 333355  & 331082  & 308297  & 246444 & 161572 & 60538  & 22691  & 11712 & 6943  & 3114  & 1662  & 162 \\

\midrule
\multicolumn{13}{@{}l}{\textbf{Top 2000}} \\
\quad All authors & 2975085 & 2952224 & 2688622 & 2061405 & 1329820 & 550527 & 230004 & 124531 & 76598 & 35735 & 19652 & 2091 \\
\quad Men         & 1321505 & 1308781 & 1182983 & 912577  & 614076  & 278944 & 125784 & 71223  & 45121 & 22047 & 12584 & 1487 \\
\quad Women       & 826410  & 821711  & 755330  & 579968  & 358068  & 125649 & 43512  & 20656  & 11606 & 4747  & 2396  & 223 \\

\bottomrule
\end{tabular}%
}
\end{table*}

\begin{figure*}[t]
\centering
\begin{tabular}{cc}
    \includegraphics[width=0.45\linewidth]{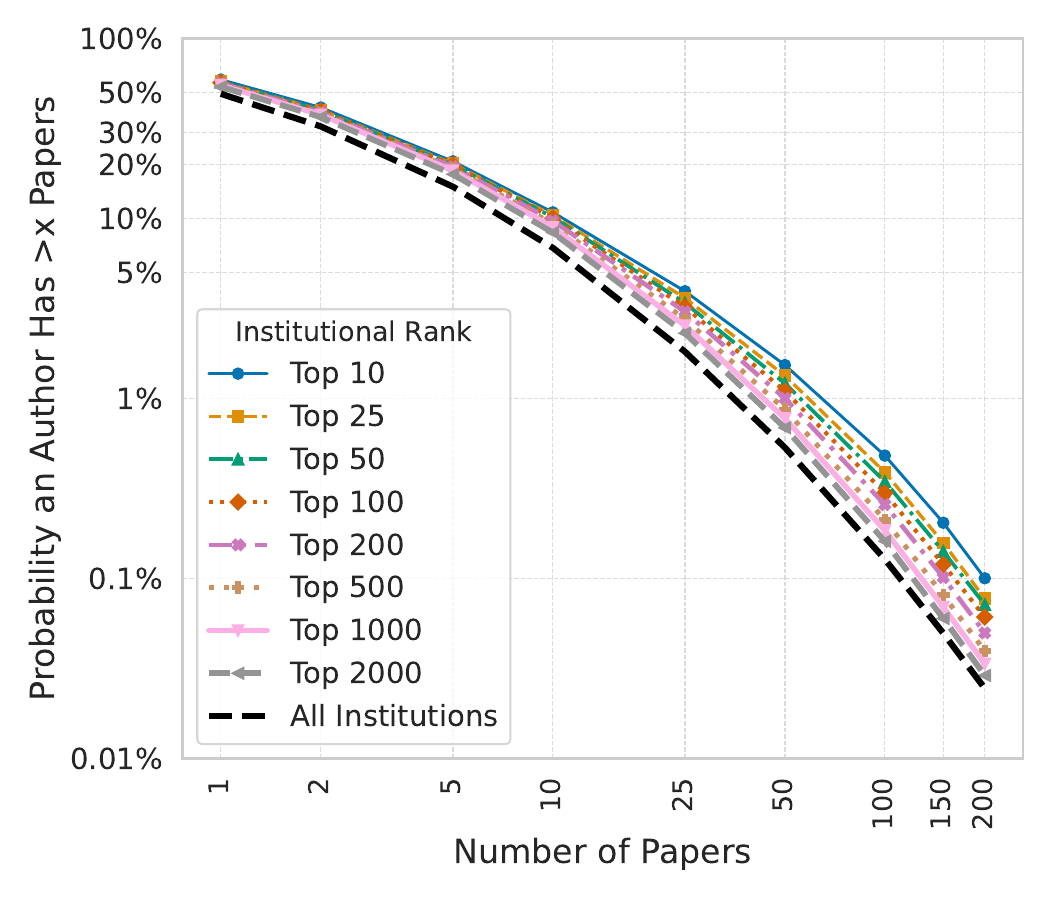}
&   
    \includegraphics[width=0.45\linewidth]{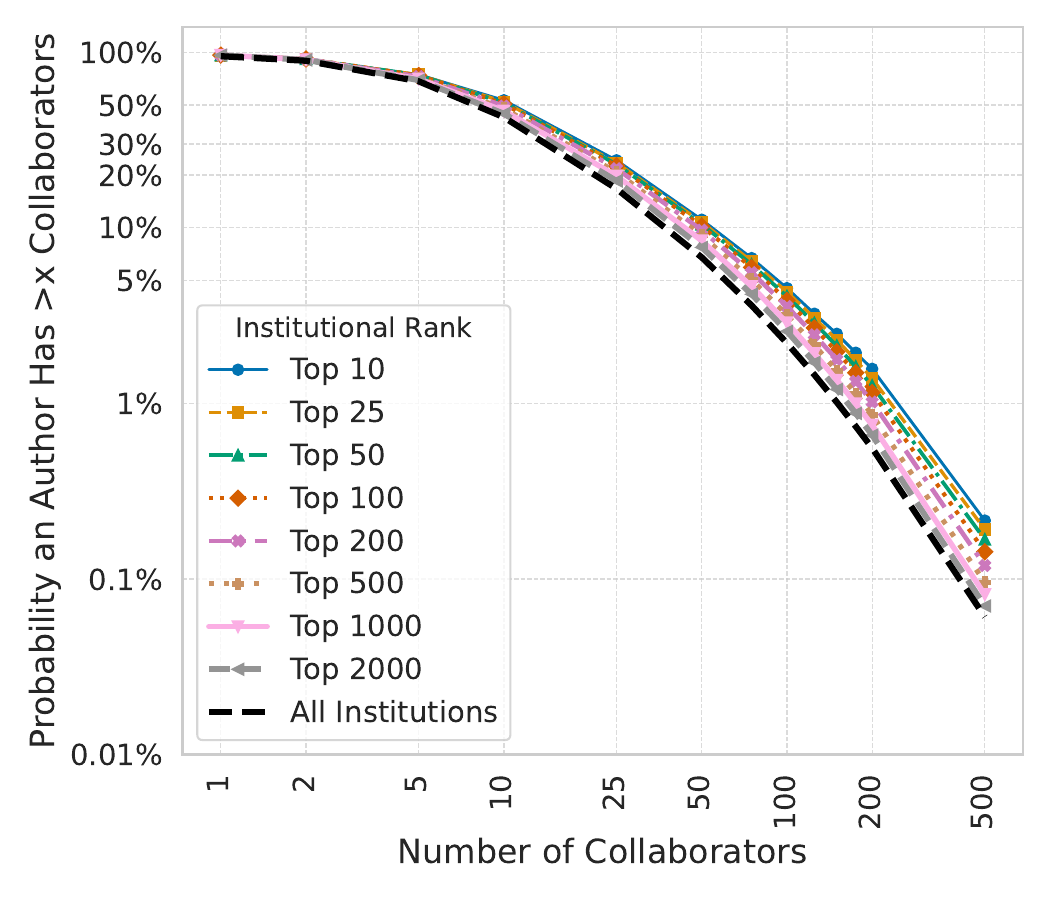} 
\\
(a) & (b) 
\end{tabular}
    \caption{\textbf{Distribution of author productivity and collaboration across institutional rank groups.}
    Complementary cumulative distribution functions (CCDFs) showing the probability that an author exceeds a given threshold $x$ of (a) publications in high-impact venues or (b) collaborators across these venues. Curves correspond to authors affiliated with institutions within different \textit{Times Higher Education} (THE) rank thresholds (top $k$ institutions), with the dashed black line indicating the distribution across all institutions. CCDFs are evaluated as $P(X>x)$ and plotted on log–log axes to highlight differences in the upper tail of the productivity and collaboration distributions.}
    \label{fig:prestige-THE-CCDF}
\end{figure*}

\end{document}